\documentclass[a4paper,11pt]{article}
\pdfoutput=1 

\usepackage{jheppub} 

\usepackage[T1]{fontenc}
\usepackage{multirow,bbold,slashed,wasysym}

\allowdisplaybreaks
 
\definecolor{nicered}{rgb}{0.7,0.1,0.1} 
\definecolor{nicegreen}{rgb}{0.1,0.5,0.1}
\definecolor{niceblue}{rgb}{0.0,0.1,0.7}
\hypersetup{colorlinks,citecolor=niceblue,linkcolor=niceblue,urlcolor=niceblue}

\usepackage[normalem]{ulem}
\usepackage[table]{xcolor}

\def \beq{\begin{equation}}
\def \eeq{\end{equation}}
\def \bea{\begin{eqnarray}}
\def \eea{\end{eqnarray}}

\title{Dijet bounds on third-generation four-quark operators}

\author[a]{Maximilian Freiheit}
\author[a]{and Ulrich Haisch}

\affiliation[a]{Max Planck Institute for Physics, \\ Boltzmannstr.~8, 85748 Garching, Germany}

\emailAdd{maximilian.freiheit@mpp.mpg.de,haisch@mpp.mpg.de}

\preprint{MPP-2025-217} 

\abstract{We use dijet measurements from the Large Hadron Collider to constrain ten third-generation four-quark operators in the Standard Model effective field theory. At tree level, only the five operators involving four bottom quarks are directly constrained, but renormalization group~(RG) effects allow all ten operators to be probed. Our analysis includes the dominant leading-logarithmic RG contributions up to two-loop order. The~resulting bounds for the first five operators are nominal stronger or comparable to current limits, while those for the remaining operators remain weak despite the inclusion of logarithmically enhanced corrections.}

\begin{document} 
\maketitle
\flushbottom

\section{Introduction} 
\label{sec:introduction}

One idea that has stood the test of time since 1985 is the concept of an effective Lagrangian for new interactions and flavor conservation~\cite{Buchmuller:1985jz}. Today, this framework is known as the Standard~Model effective field theory~(SMEFT)~\cite{Grzadkowski:2010es,Brivio:2017vri,Isidori:2023pyp} and has become a central tool for probing indirect signatures of physics beyond the Standard Model~(BSM) at the Large Hadron Collider~(LHC). The increasing precision of experimental measurements calls for equally precise theoretical predictions, both within the Standard Model~(SM) and for possible BSM effects.

A notable feature of the SMEFT --- sometimes viewed as a bug, sometimes as a feature --- is the large number of effective operators and their mixing under the renormalization group~(RG) flow~\cite{Jenkins:2013zja,Jenkins:2013wua,Alonso:2013hga,Gorbahn:2016uoy,Bern:2020ikv,Jin:2020pwh,Haisch:2022nwz,DiNoi:2023ygk,Jenkins:2023bls,DiNoi:2024ajj,Born:2024mgz,Naterop:2024cfx,Duhr:2025zqw,Haisch:2025lvd,Zhang:2025ywe,Assi:2025fsm,Naterop:2025cwg,DiNoi:2025arz,Haisch:2025vqj,Duhr:2025yor,DiNoi:2025tka,Banik:2025wpi,Henriksson:2025vyi,Chala:2025crd,Guedes:2025sax}. Perhaps the most important consequence of the RG evolution is that it inevitably breaks the accidental symmetries of the SM, such as the flavor $U(3)^5$ or the custodial $SU(2)$ symmetry. As a result, operators that are only weakly constrained by direct tree-level measurements can, in some cases, be probed indirectly through loop-level observables. A~prime example is the purely right-handed top four-quark operator, which induces two-loop double-logarithmic contributions to the $\rho$ parameter~\cite{Allwicher:2023aql,Allwicher:2023shc,Stefanek:2024kds,Haisch:2024wnw}, a key quantity in electroweak~(EW) precision measurements. Similar two-loop double-logarithmic effects also arise in the processes $gg \to h$ and $h \to \gamma \gamma$~\cite{Haisch:2025lvd,Haisch:2025vqj}, although they are phenomenologically less significant. In this article, we emphasize that two-loop double-logarithmic contributions can mediate mixing from third-generation to first- and second-generation four-quark operators. This effect may be important, as the Wilson~coefficients of four-quark operators involving third-generation fermions remain only weakly constrained~\cite{Gauld:2015lmb,Ethier:2021bye,Dawson:2022bxd,Aoude:2022deh,ATLAS:2023ajo,Degrande:2024mbg,Haisch:2024wnw,Haisch:2025vqj,DiNoi:2025uhu}. Consequently, once radiative corrections from RG evolution are included, dijet measurements at the LHC could provide meaningful constraints on these third-generation four-quark operators. Although jet physics has been employed to constrain SMEFT operators~\cite{Krauss:2016ely,Alioli:2017jdo,Alte:2017pme,Hirschi:2018etq,Keilmann:2019cbp,Goldouzian:2020wdq,Haisch:2021hcg,Degrande:2025vhl}, to~our knowledge, constraints specifically on third-generation four-quark operators have not yet been established.

The remainder of this article is organized as follows. In Section~\ref{sec:framework}, we introduce the subset of dimension-six SMEFT operators relevant to our analysis. Section~\ref{sec:calculation} exemplifies the main steps in computing the SMEFT corrections to dijet production. The phenomenological implications of our calculations are discussed in Section~\ref{sec:phenomenology}, and we conclude in Section~\ref{sec:conclusions}. A comparison of the constraints on the third-generation four-quark operators derived here with those in the literature is given in Appendix~\ref{app:comparison}.

\section{Theoretical framework} 
\label{sec:framework}

To establish our notation and conventions, we begin by defining the SMEFT Lagrangian as
\beq \label{eq:LSMEFT}
{\cal L}_{\rm SMEFT} = \sum_i C_i(\mu_R)\, Q_i \,.
\eeq
Here, $C_i(\mu_R)$ denotes the dimensionful Wilson~coefficients evaluated at the renormalization scale $\mu_R$, which multiply the corresponding effective operators $Q_i$. Throughout this work, we assume that all Wilson~coefficients are real.

In the Warsaw basis~\cite{Grzadkowski:2010es}, the dimension-six four-quark operators relevant to this work are given by:
\begin{gather}
Q^{(1)}_{qq, ijkl} = (\bar q_i \gamma_\mu q_j) (\bar q_k \gamma^\mu q_l) \,, \qquad Q^{(3)}_{qq, ijkl} = (\bar q_i \gamma_\mu \sigma^I q_j) (\bar q_k \gamma^\mu \sigma^I q_l) \,, \nonumber \\[2mm]
Q_{uu, ijkl} = (\bar u_i \gamma_\mu u_j) (\bar u_k \gamma^\mu u_l) \,, \qquad Q_{dd, ijkl} = (\bar d_i \gamma_\mu d_j) (\bar d_k \gamma^\mu d_l) \,, \nonumber \\[2mm]
Q^{(1)}_{ud, ijkl} = (\bar u_i \gamma_\mu u_j) (\bar d_k \gamma^\mu d_l) \,, \qquad Q^{(8)}_{ud, ijkl} = (\bar u_i \gamma_\mu T^A u_j) (\bar d_k \gamma^\mu T^A d_l) \,, \label{eq:4Foperators} \\[2mm]
Q^{(1)}_{qu, ijkl} = (\bar q_i \gamma_\mu q_j) (\bar u_k \gamma^\mu u_l) \,, \qquad Q^{(8)}_{qu, ijkl} = (\bar q_i \gamma_\mu T^A q_j)(\bar u_k \gamma^\mu T^A u_l) \,, \nonumber \\[2mm]
Q^{(1)}_{qd, ijkl} = (\bar q_i \gamma_\mu q_j) (\bar d_k \gamma^\mu d_l) \,, \qquad Q^{(8)}_{qd, ijkl} = (\bar q_i \gamma_\mu T^A q_j)(\bar d_k \gamma^\mu T^A d_l) \,. \nonumber
\end{gather}
The symbol $q$ denotes the left-handed quark $SU(2)_L$ doublets, while $u$ and $d$ denote the right-handed up- and down-quark $SU(2)_L$ singlets, respectively. The indices $i$, $j$, $k$, and~$l$ represent flavor indices in the weak eigenstate basis. The symbols $\sigma^I$ denote the Pauli matrices, and $T^A = \lambda^A/2$ are the $SU(3)_C$ generators, with $\lambda^A$ representing the Gell-Mann matrices. Of particular interest in what follows are the third-generation four-quark operators, i.e., the operators in~(\ref{eq:4Foperators}) with $i = j = k = l = 3$. These operators will be denoted by $Q^{(1)}_{QQ}$, $Q^{(3)}_{QQ}$, $Q_{tt}$, $Q_{bb}$, $Q^{(1)}_{tb}$, $Q^{(8)}_{tb}$, $Q^{(1)}_{Qt}$, $Q^{(8)}_{Qt}$, $Q^{(1)}_{Qb}$, and $Q^{(8)}_{Qb}$. Before proceeding, we note that in the phenomenological analysis of Section~\ref{sec:phenomenology}, all quark Yukawa couplings except that of the top quark are set to zero. Moreover, the Cabibbo-Kobayashi-Maskawa matrix is taken to be diagonal when transforming the quark fields from the weak to the mass eigenstate basis.

\section{Anatomy of dijet constraints} 
\label{sec:calculation}

In this section, we outline the essential ingredients and calculations needed to use searches for light jet pairs to constrain the Wilson~coefficients of the third-generation four-quark operators introduced in~\eqref{eq:4Foperators}. The observable of interest is the dijet angular distribution,~i.e., the differential cross section for a pair of jets with invariant mass $M_{jj}$ produced at an angle~$\hat{\theta}$ relative to the beam direction in the jet-jet center-of-mass~(CM) frame. 

Unlike conventional resonance searches, the dijet angular distribution provides the advantage of constraining broad $s$-channel resonances or modifications of the spectrum arising from higher-dimensional operators in a relatively model-independent manner. This is because the dominant channels in QCD dijet production follow the familiar Rutherford scattering behavior, 
\beq \label{eq:rutherford}
\frac{d\sigma}{d\cos\hat{\theta}} \propto \frac{1}{\sin^4(\hat{\theta}/2)} \,,
\eeq
at small angles $\hat{\theta}$, characteristic of $t$-channel exchange of a massless spin-$1$ boson. To~regularize the Rutherford singularity, one commonly studies dijet cross sections differential~in
\beq \label{eq:chi}
\chi = \frac{1 + \cos\hat{\theta}}{1 - \cos\hat{\theta}} \,.
\eeq
In the small-angle limit, $\chi \to \infty$, the partonic differential QCD cross section becomes approximately constant,
\beq \label{eq:chirutherford}
\frac{d\sigma}{d\chi} \propto \text{const.} 
\eeq
The presence of a heavy resonance or an effective operator can induce additional hard scatterings, resulting in increased jet production perpendicular to the beam. As a consequence, one expects deviations from the QCD prediction, manifested as enhanced high-energy jet activity in the central region of the detector. In particular, if the angular distributions are influenced by a heavy degree of freedom or an effective operator, an excess of events is expected in $d\sigma/d\chi$ for $\chi \to 1$ at large $M_{jj}$, relative to the approximately flat QCD background. In contrast, heavy BSM physics is not expected to leave a significant imprint in the $d\sigma/d\chi$ distribution for $\chi \to \infty$.

\begin{figure}[t!]
\begin{center}
\includegraphics[width=0.85\textwidth]{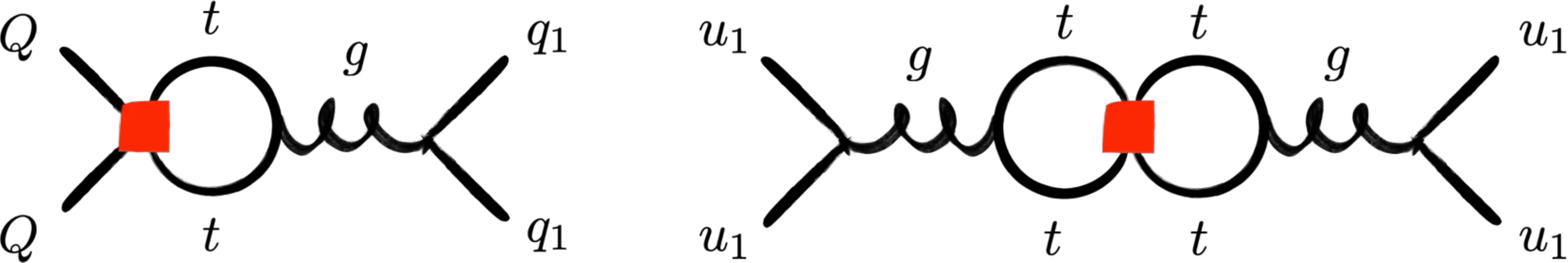} 
\end{center}
\vspace{0mm} 
\caption{Left: Example of a QCD penguin diagram. Right: Example of a QCD double-penguin diagram. The red boxes indicate operator insertions. The leading UV poles of these diagrams determine, respectively, the one-loop logarithmic and two-loop double-logarithmic corrections in the SMEFT RG flow of the third-generation four-quark operators. See main~text for more details. \label{fig:diagrams}} 
\end{figure}

The above discussion is best illustrated with simple examples relevant to this analysis. We first note that, owing to the proton's nonzero bottom-quark parton distribution function~(PDF) in the five-flavor scheme, all third-generation four-quark operators with four bottom quarks contribute at tree level to dijet production at the LHC. In the notation introduced in the previous section, these are the five operators $Q^{(1)}_{QQ}$, $Q^{(3)}_{QQ}$, $Q_{bb}$, $Q^{(1)}_{Qb}$, and~$Q^{(8)}_{Qb}$. In the case of the operator $Q^{(8)}_{Qb}$, and assuming massless bottom quarks, the squared matrix element for tree-level $bb \to bb$ scattering, including the pure SM contribution, is, for example, given by
\beq \label{eq:treeexample}
\begin{split}
\overline{\sum} \left | {\cal M} ( b b \to b b ) \right |^2 & = \frac{256 \hspace{0.125mm} \pi^2 \hspace{0.125mm} \alpha_s^2 (M_{jj}) \left(3 \chi^4 + 2 \chi^3 + \chi^2 + 2 \chi + 3\right)}{27 \hspace{0.125mm} \chi^2 \left (\chi + 1 \right )^2} \\[2mm]
& \phantom{xx} - \frac{64 \hspace{0.125mm} \pi \hspace{0.125mm} \alpha_s (M_{jj}) M_{jj}^2 \left ( \chi^2 - \chi + 1 \right )}{9 \hspace{0.125mm} \chi \left ( \chi + 1 \right )^2} \, C^{(8)}_{Qb} (M_{jj}) \\[2mm]
& \phantom{xx} + \frac{8 M_{jj}^4 \left ( \chi^2 + 1 \right )}{9 \left (\chi + 1 \right )^4} \, \big ( C^{(8)}_{Qb} (M_{jj}) \big )^2 \,.
\end{split}
\eeq
Here, the symbol $\overline{\sum}$ indicates that the color and spin indices are averaged over the initial states and summed over the final states, while $\alpha_s = g_s^2/(4 \pi)$ denotes the QCD coupling constant. As seen from~\eqref{eq:treeexample}, the terms linear and quadratic in $C^{(8)}_{Qb}$ are enhanced by two and four powers of the dijet invariant mass, respectively, relative to the SM contribution, reflecting the higher-dimensional nature of the effective interactions introduced in~\eqref{eq:4Foperators}. The~renormalization scale in both $\alpha_s$ and $C^{(8)}_{Qb}$ has been set to $M_{jj}$ in~\eqref{eq:treeexample}, which represents the natural choice. Apart from $Q^{(1)}_{Qb}$, which does not interfere with the SM due to its color structure, the squared matrix elements of $Q^{(1)}_{QQ}$, $Q^{(3)}_{QQ}$, and $Q_{bb}$ have the same structure as above. Similar considerations apply to $b \bar b \to b \bar b$ scattering. 

Since the five-flavor scheme lacks a top-quark PDF and dijet searches do not involve top quarks in the final state, the top-quark counterpart $Q^{(8)}_{Qt}$ of the operator~$Q^{(8)}_{Qb}$ does not contribute to dijet production at tree level. However, if matching to a ultraviolet~(UV) complete model at the scale $\Lambda$ generates a nonzero Wilson~coefficient $C^{(8)}_{Qt}(\Lambda)$, the one-loop RG flow in the SMEFT will induce, at the scale $M_{jj}$, nonzero Wilson~coefficients for four-quark operators with first- or second-generation content that contribute to dijet production at tree level. At leading-logarithmic~(LL) accuracy, the relevant one-loop corrections are
\begin{gather}
\frac{C^{(1)}_{qq, ii33} (M_{jj})}{C^{(8)}_{Qt}(\Lambda)} \simeq \frac{\alpha_s (\Lambda)}{72 \pi} \hspace{0.25mm} L \,, \quad \frac{C^{(1)}_{qq, i33i} (M_{jj})}{C^{(8)}_{Qt}(\Lambda)} \simeq -\frac{\alpha_s (\Lambda)}{48 \pi} \hspace{0.25mm} L \,, \quad \frac{C^{(3)}_{qq, i33i} (M_{jj})}{C^{(8)}_{Qt}(\Lambda)} \simeq -\frac{\alpha_s (\Lambda)}{48 \pi} \hspace{0.25mm} L \,, \nonumber \\[-3mm] \label{eq:oneloopLL} \\[-3mm]
\frac{C^{(8)}_{qu, 33ii} (M_{jj})}{C^{(8)}_{Qt}(\Lambda)} \simeq -\frac{\alpha_s (\Lambda)}{6 \pi} \hspace{0.25mm} L \,, \quad \frac{C^{(8)}_{qd, 33ii} (M_{jj})}{C^{(8)}_{Qt}(\Lambda)} \simeq -\frac{\alpha_s (\Lambda)}{6 \pi} \hspace{0.25mm} L \,, \nonumber 
\end{gather}
where $i = 1, 2$ and we have introduced the abbreviation $L = \ln \left ( \Lambda/M_{jj} \right )$. The results in~\eqref{eq:oneloopLL} can be derived from the SMEFT anomalous dimensions computed in~\cite{Alonso:2013hga}. In the language of flavor physics, the relevant anomalous dimensions result from QCD penguin insertions. An example diagram is shown on the left-hand side of~Figure~\ref{fig:diagrams}. The~corrections in~(\ref{eq:oneloopLL}) therefore induce logarithmically enhanced one-loop contributions to the squared matrix element relevant for dijet production. For instance, in the case of $b d \to b d$ scattering, the squared matrix element resulting from~(\ref{eq:oneloopLL}), including the SM contribution, takes the form:
\beq \label{eq:oneexample}
\begin{split}
\overline{\sum} \left | {\cal M} ( b d \to b d ) \right |^2 & \simeq \frac{128 \hspace{0.125mm} \pi^2 \hspace{0.125mm} \alpha_s^2 (M_{jj}) \left(2 \chi^2 + 2 \chi + 1\right)}{9 \hspace{0.125mm} \left (\chi + 1 \right )^2} \\[2mm]
& \phantom{xx} + \frac{16 \hspace{0.25mm} \hspace{0.125mm} \alpha_s (M_{jj}) \hspace{0.25mm} \alpha_s (\Lambda) M_{jj}^2 \left ( 2 \chi^2 + 2 \chi + 1 \right )}{27 \hspace{0.125mm}\left ( \chi + 1 \right )^3} \, L \hspace{0.25mm} C^{(8)}_{Qt} (\Lambda) \\[2mm]
& \phantom{xx} + \frac{\alpha_s^2 (\Lambda) M_{jj}^4 \left ( 2 \chi^2 + 2 \chi + 1 \right )}{81 \pi^2 \left (\chi + 1 \right )^4} \, \big ( L \hspace{0.25mm} C^{(8)}_{Qt} (\Lambda) \big )^2 \,.
\end{split}
\eeq 
Similar considerations apply to the other relevant partonic $2 \to 2$ scattering processes.

Unlike the third-generation four-quark operators $Q^{(1)}_{QQ}$, $Q^{(3)}_{QQ}$, $Q_{bb}$, $Q^{(1)}_{tb}$, $Q^{(8)}_{tb}$, $Q^{(1)}_{Qt}$, $Q^{(8)}_{Qt}$, $Q^{(1)}_{Qb}$, and $Q^{(8)}_{Qb}$, the purely right-handed top four-quark operator $Q_{tt}$ does not induce logarithmically enhanced one-loop corrections in dijet production. This is intuitively clear, since converting two top-quark currents into two light-quark currents requires a double-penguin operator insertion. The right side of~Figure~\ref{fig:diagrams} presents an example of such a diagram. As~a~result of this two-step mixing process, one, however, obtains double-logarithmic two-loop corrections. See, for instance,~\cite{Buras:2018gto,Haisch:2025lvd} for recent discussions, including a resummation of such effects. At LL accuracy, the relevant two-loop corrections inducing purely first- or second-generation four-quark operators are
\begin{gather}
\frac{C^{(1)}_{qq, iiii} (M_{jj})}{C_{tt}(\Lambda)} \simeq \frac{9 \hspace{0.125mm} \alpha_s^2 (\Lambda) + 32 \hspace{0.125mm} \alpha_1^2 (\Lambda)}{1944 \pi^2} \hspace{0.25mm} L^2 \,, \quad \frac{C^{(3)}_{qq, iiii} (M_{jj})}{C_{tt}(\Lambda)} \simeq \frac{\alpha_s^2 (\Lambda)}{72 \pi^2} \hspace{0.25mm} L^2 \,, \nonumber \\[2mm] 
\frac{C_{uu, iiii} (M_{jj})}{C_{tt}(\Lambda)} \simeq \frac{9 \hspace{0.125mm} \alpha_s^2 (\Lambda) + 128 \hspace{0.125mm} \alpha_1^2 (\Lambda)}{486 \pi^2} \hspace{0.25mm} L^2 \,, \quad \frac{C_{dd, iiii} (M_{jj})}{C_{tt}(\Lambda)} \simeq \frac{9 \alpha_s^2 (\Lambda) + 32 \alpha_1^2 (\Lambda)}{486 \pi^2} \hspace{0.25mm} L^2 \,, \nonumber \\[2mm]
\frac{C^{(1)}_{qu, iiii} (M_{jj})}{C_{tt}(\Lambda)} \simeq \frac{32 \hspace{0.125mm} \alpha_1^2 (\Lambda)}{243 \pi^2} \hspace{0.25mm} L^2 \,, \quad \frac{C^{(8)}_{qu, iiii} (M_{jj})}{C_{tt}(\Lambda)} \simeq \frac{\alpha_s^2 (\Lambda)}{9 \pi^2} \hspace{0.25mm} L^2 \,, \label{eq:twoloopLL} \\[2mm] 
\frac{C^{(1)}_{qd, iiii} (M_{jj})}{C_{tt}(\Lambda)} \simeq -\frac{16 \hspace{0.125mm} \alpha_1^2 (\Lambda)}{243 \pi^2} \hspace{0.25mm} L^2 \,, \quad \frac{C^{(8)}_{qd, iiii} (M_{jj})}{C_{tt}(\Lambda)} \simeq \frac{\alpha_s^2 (\Lambda)}{9 \pi^2} \hspace{0.25mm} L^2 \,,\nonumber \\[2mm] 
\frac{C^{(1)}_{ud, iiii} (M_{jj})}{C_{tt}(\Lambda)} \simeq -\frac{64 \hspace{0.125mm} \alpha_1^2 (\Lambda)}{243 \pi^2} \hspace{0.25mm} L^2 \,, \quad \frac{C^{(8)}_{ud, iiii} (M_{jj})}{C_{tt}(\Lambda)} \simeq \frac{\alpha_s^2 (\Lambda)}{9 \pi^2} \hspace{0.25mm} L^2 \,. \nonumber 
\end{gather}
Here, again, $i = 1,2$, and we have introduced $\alpha_1 = g_1^2/(4\pi)$, with $g_1$ denoting the $U(1)_Y$ gauge coupling. It is relatively straightforward to derive the results~(\ref{eq:twoloopLL}) from the SMEFT anomalous dimensions computed in~\cite{Alonso:2013hga}. Notice that some of the Wilson~coefficients at~$M_{jj}$ are induced by QED double-penguin insertions, due to the color structure of the corresponding operators. Using~(\ref{eq:twoloopLL}), one can compute the induced LL-enhanced two-loop contributions to the squared matrix element relevant for dijet production. In the case of $u u \to u u$ scattering, we find, for example, the following expression for the full SM~plus~BSM contribution arising from a nonzero initial condition $C_{tt} (\Lambda)$:
\beq \label{eq:twoexample}
\begin{split}
\overline{\sum} \left | {\cal M} ( u u \to u u ) \right |^2 & \simeq \frac{256 \hspace{0.125mm} \pi^2 \hspace{0.125mm} \alpha_s^2 (M_{jj}) \left(3 \chi^4 + 2 \chi^3 + \chi^2 + 2 \chi + 3\right)}{27 \hspace{0.125mm} \chi^2 \left (\chi + 1 \right )^2} \\[2mm]
& \phantom{xx} - \frac{64 \hspace{0.25mm} \hspace{0.125mm} \alpha_s (M_{jj}) \hspace{0.25mm} \alpha_s^2 (\Lambda) M_{jj}^2 \left ( 5 \chi^2 + \chi + 5 \right )}{243 \pi \hspace{0.125mm} \chi \left ( \chi + 1 \right )^2} \, L^2 \hspace{0.25mm} C_{tt} (\Lambda) \\[2mm]
& \phantom{xx} + \frac{8 \alpha_s^4 (\Lambda) M_{jj}^4 \left ( 7 \chi^2 + 8 \chi + 7 \right )}{2187 \pi^4 \left (\chi + 1 \right )^4} \, \big ( L^2 \hspace{0.25mm} C_{tt} (\Lambda) \big )^2 \,.
\end{split}
\eeq 
Here, for simplicity, we have dropped subleading terms involving powers of $\alpha_1$. The structure of the squared matrix elements for the other relevant $2 \to 2$ channels is similar to that shown in~(\ref{eq:twoexample}).

The previous discussion has shown that including RG effects induces logarithmically enhanced loop contributions to dijet production associated with third-generation four-quark operators that do not contribute at tree level. To assess whether this observation can be phenomenologically relevant, it is important to recall that, in order to calculate the differential dijet cross sections, one requires not only the squared matrix elements but also the parton-parton luminosities, defined as
\beq \label{eq:lumis}
f\hspace{-1.3mm}f_{ij} \left (\tau, \mu_F \right ) = \frac{2}{1 + \delta_{ij}} \int_{\tau}^{1} \! \frac{dx}{x} \, f_{i/p} \left (x, \mu_F \right ) \, f_{j/p} \left ( \frac{\tau}{x} , \mu_F \right ) \,. 
\eeq
Here, $\tau = M_{jj}^2/s$, where $s$ is the CM energy of the collider, and $f_{i/p}(x, \mu_F)$ denotes the universal, non-perturbative PDF describing the probability of finding parton $i$ inside the proton~($p$) with longitudinal momentum fraction $x$, while $\mu_F$ is the factorization scale. Numerically, we find $f\hspace{-1.3mm}f_{bd}/f\hspace{-1.3mm}f_{bb} \in [2 \cdot 10^2, 1 \cdot 10^3]$ and $f\hspace{-1.3mm}f_{uu}/f\hspace{-1.3mm}f_{bb} \in [3 \cdot 10^3, 2 \cdot 10^5]$ for the relevant $\tau$~values. These numbers should be compared to the suppression factor of about $\alpha_s/(4 \pi) \hspace{0.25mm} L \simeq 7 \cdot 10^{-3}$ per loop, suggesting that the PDF enhancement of the quantum corrections~(\ref{eq:oneexample}) and~(\ref{eq:twoexample}) could at least partially overcome their loop suppression. To~assess the effectiveness of this enhancement mechanism, however, a more detailed study is required. This will be carried out in the next section.

\section{Phenomenology} 
\label{sec:phenomenology}

To constrain the Wilson~coefficients of the third-generation four-quark operators $Q^{(1)}_{QQ}$, $Q^{(3)}_{QQ}$, $Q_{tt}$, $Q_{bb}$, $Q^{(1)}_{tb}$, $Q^{(8)}_{tb}$, $Q^{(1)}_{Qt}$, $Q^{(8)}_{Qt}$, $Q^{(1)}_{Qb}$, and $Q^{(8)}_{Qb}$, we make use of the CMS dijet analysis~\cite{CMS:2018ucw}. This analysis is based on an integrated luminosity of $36 \, {\rm fb}^{-1}$ collected during LHC~Run~2 and has also been used in the previous SMEFT studies~\cite{Haisch:2021hcg,Degrande:2025vhl}. It presents dijet angular distributions normalized to unity across seven $M_{jj}$ intervals, reported as $1/\sigma \hspace{0.25mm} d\sigma/d\chi$, where~$\sigma$~denotes the dijet cross section within the analysis phase space. We note that more recent differential dijet measurements using the full LHC~Run~2 dataset of $140 \, {\rm fb}^{-1}$ have been performed by both the ATLAS and CMS collaborations~\cite{CMS:2023fix,ATLAS:2025ifq}. Unfortunately, both analyses report $M_{jj}$ distributions binned in $y^\ast$, rather than $\chi$ distributions binned in $M_{jj}$, with $y^\ast = 1/2 \hspace{0.125mm} \ln \chi$. Using the results presented in~\cite{CMS:2023fix,ATLAS:2025ifq} in the context of our work would therefore be more involved, and we refrain from doing so.

\begin{figure}[t!]
\begin{center}
\includegraphics[width=0.475\textwidth]{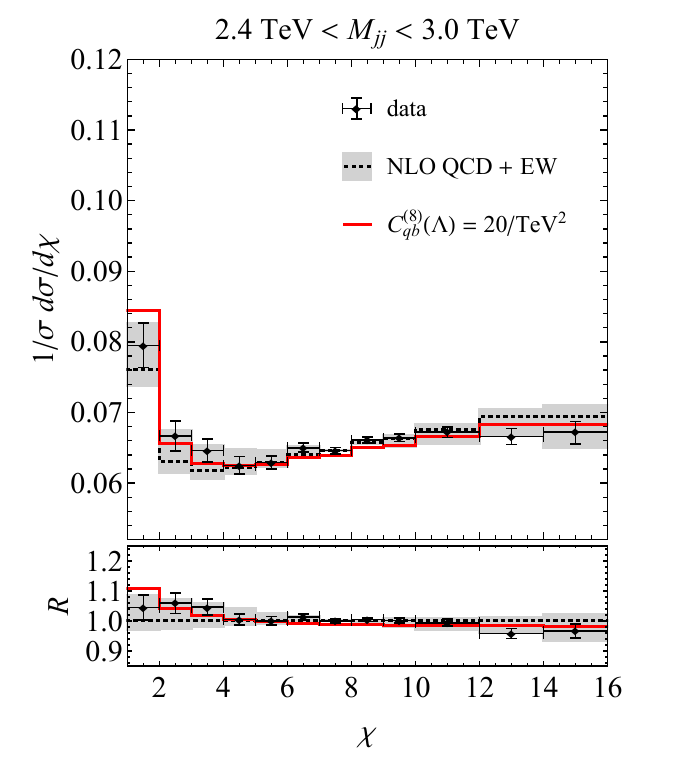} \quad 
\includegraphics[width=0.475\textwidth]{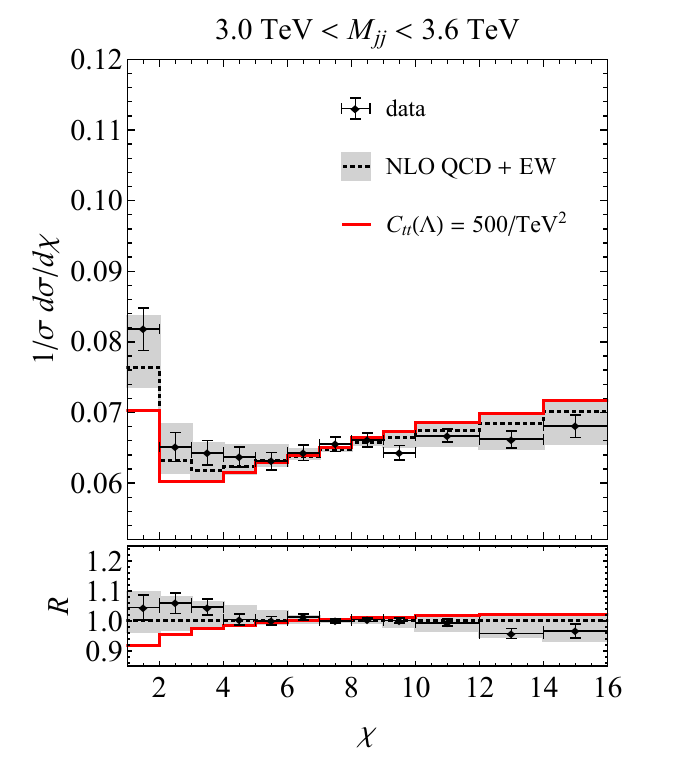}
\end{center}
\vspace{-4mm} 
\caption{Normalized $\chi$ distributions in the two lowest mass bins considered in the CMS analysis~\cite{CMS:2018ucw}. The SM predictions obtained by CMS, including NLO QCD and EW corrections, are shown as black dotted lines. Error bars denote the statistical and experimental systematic uncertainties added in quadrature, while the gray band represents the theoretical uncertainties. For comparison, the normalized dijet angular distributions for $C^{(8)}_{Qb}(\Lambda) = 20/{\rm TeV}^2$ and $C_{tt}(\Lambda) = 500/{\rm TeV}^2$ are shown as red lines in the left- and right-hand panels, respectively. The lower panels display the ratios~($R$) of the unfolded data to the SM predictions, together with the corresponding BSM distributions. Additional explanations can be found in the main text. \label{fig:bins}} 
\end{figure}

Our dijet predictions are based on the {\tt CT14nnlo\_as\_0118} PDFs~\cite{Dulat:2015mca}, with both the renormalization and factorization scales, $\mu_R$ and $\mu_F$, chosen to be $M_{jj}$. The same PDF set was used in~\cite{CMS:2018ucw} and is implemented in our calculations via {\tt ManeParse}~\cite{Clark:2016jgm}. The normalized $1/\sigma \hspace{0.25mm} d\sigma/d\chi$ distributions are rescaled by bin-dependent multiplicative factors, defined as the ratio of the SM predictions provided by the CMS collaboration to our leading order~(LO) SM results. The CMS SM predictions incorporate next-to-leading order (NLO) QCD corrections computed with {\tt NLOJET++}~\cite{Nagy:2001fj}, as well as the NLO EW corrections evaluated in~\cite{Dittmaier:2012kx}. Non-perturbative effects arising from hadronization and multiple parton interactions are investigated in~\cite{CMS:2018ucw} using {\tt PYTHIA~8}~\cite{Sjostrand:2007gs} and {\tt HERWIG++}~\cite{Bahr:2008pv}. For both Monte Carlo~(MC) generators, these effects are found to be below the $1\%$ level and therefore negligible.  Since for low values of $\chi$ --- the region most relevant for the analysis presented below --- next-to-next-to-leading order~(NNLO) QCD corrections to dijet production are only moderate, at the level of~$5\%$ or below~\cite{Chen:2022tpk,NNLOJET:2025rno}, we neglect NNLO QCD effects in our article. The resulting multiplicative factors lie in the range of about $[0.95, 1.30]$, demonstrating good agreement between our LO SM predictions and the more complete SM calculations performed by CMS. These factors are subsequently applied to both the SM and BSM predictions, thereby effectively upgrading them to NLO QCD and EW accuracy. This approach assumes that the size of the perturbative corrections is independent of the specific features of the BSM signal. While this assumption is well motivated by factorization arguments, a definitive validation would require a fully realistic MC simulation, which lies beyond the scope of the present~work.

In~Figure~\ref{fig:bins}, we present the $1/\sigma \hspace{0.25mm} d\sigma/d\chi$ distributions for the two lowest $M_{jj}$ regions considered by the CMS collaboration in~\cite{CMS:2018ucw}, namely $2.4 \, {\rm TeV} < M_{jj} < 3.0 \, {\rm TeV}$ and $3.0 \, {\rm TeV} < M_{jj} < 3.6 \, {\rm TeV}$. Unfolded data are compared to the SM prediction including NLO QCD and EW corrections (black dotted lines), as computed by the CMS collaboration and discussed above. The error bars represent the combined statistical and experimental systematic uncertainties, while theoretical uncertainties are shown as gray bands. For comparison, the normalized dijet angular distributions assuming $C^{(8)}_{Qb} (\Lambda) = 20/{\rm TeV}^2$ and $C_{tt} (\Lambda) = 500/{\rm TeV}^2$ are also shown on the left- and right-hand sides, respectively, as red lines.  The initial scale is chosen as $\Lambda = 10 \, {\rm TeV}$, and all initial conditions of the third-generation four-quark operators, except for the one indicated in the panel, are set to zero. The RG evolution from $\Lambda$ to $M_{jj}$ is performed keeping all LL terms up to two loops, including the full dependence on the three SM gauge couplings and the top-quark Yukawa coupling. Both panels clearly show that, as anticipated, the largest relative deviations between the BSM and SM results occur in the lowest bin, where $\chi \in [1, 2]$. Note also that for~$C^{(8)}_{Qb}(\Lambda) = 20/{\rm TeV}^2$, the BSM contribution is constructive, as it is dominated by the term quadratic in $C^{(8)}_{Qb}$. In contrast, for $C_{tt}(\Lambda) = 500/{\rm TeV}^2$, one observes destructive interference arising from the term linear~in~$C_{tt}$. We note that the values of $C^{(8)}_{Qb}(\Lambda)$ and $C_{tt}(\Lambda)$ used to obtain the BSM predictions in the figure exceed the perturbativity limit. Consequently, they are primarily intended for illustrative purposes, a caveat that should be kept in mind when interpreting these benchmarks in the context of explicit BSM constructions.

We extract the 95\% confidence-level (CL) limits on the Wilson~coefficients of the third-generation four-quark operators by performing a $\chi^2$~fit to the three lowest $M_{jj}$ bins reported by CMS in~\cite{CMS:2018ucw}, spanning $2.4 \, {\rm TeV} < M_{jj} < 4.2 \, {\rm TeV}$. These bins are chosen because they provide the highest sensitivity to the effects under study once both experimental and theoretical uncertainties are incorporated. Although the BSM predictions grow with energy, the statistical uncertainties of the data increase noticeably for $M_{jj} > 4.2 \, {\rm TeV}$, rendering the higher $M_{jj}$ bins largely ineffective for constraining the Wilson~coefficients. In~the~$\chi^2$~analysis, all other Wilson~coefficients are treated as nuisance parameters and profiled over, thereby accounting for correlations among operators. This procedure yields conservative and robust bounds without assuming that all other coefficients vanish. We have also explicitly checked that profiling produces slightly weaker constraints than marginalization.

{
\renewcommand{\arraystretch}{1.25}
\begin{table}[t!]
\centering
\begin{tabular}{|c|c|c|c|c|}
\hline
$C^{(1)}_{QQ} (v)$ & $C^{(3)}_{QQ} (v)$ & $C_{tt} (v)$ & $C_{bb} (v)$ & $C^{(1)}_{Qt} (v)$ \\ \hline 
$[-68, 65]$ &$[-58, 61]$ & $[-850, 330]$ & $[-4.7, 4.7]$ & $[-290, 300]$ \\ \hline \hline
$C^{(8)}_{Qt} (v)$ & $C^{(1)}_{Qb} (v)$ & $C^{(8)}_{Qb} (v)$ & $C^{(1)}_{tb} (v)$ & $C^{(8)}_{tb} (v)$ \\ \hline 
$[-490, 500]$ &$[-6.2, 6.2]$ & $[-9.3, 9.4]$ & $[-310, 310]$ & $[-130, 170]$ \\ \hline 
\end{tabular}
\vspace{4mm} 
\caption{Individual $95\%$~CL limits on the third-generation four-quark operators, given in units of $1/{\rm TeV}^2$, derived from the three lowest $M_{jj}$ bins in the CMS analysis~\cite{CMS:2018ucw} of normalized dijet angular distributions. These~limits are obtained from a profiled $\Delta \chi^2$ analysis and correspond to the Wilson~coefficients evaluated at the scale $v = 250 \, {\rm GeV}$. Consult the main text for further details. \label{tab:1DCL2}}
\end{table}
}

\begin{figure}[t!]
\begin{center}
\includegraphics[width=\textwidth, trim = 80 0 80 0, clip]{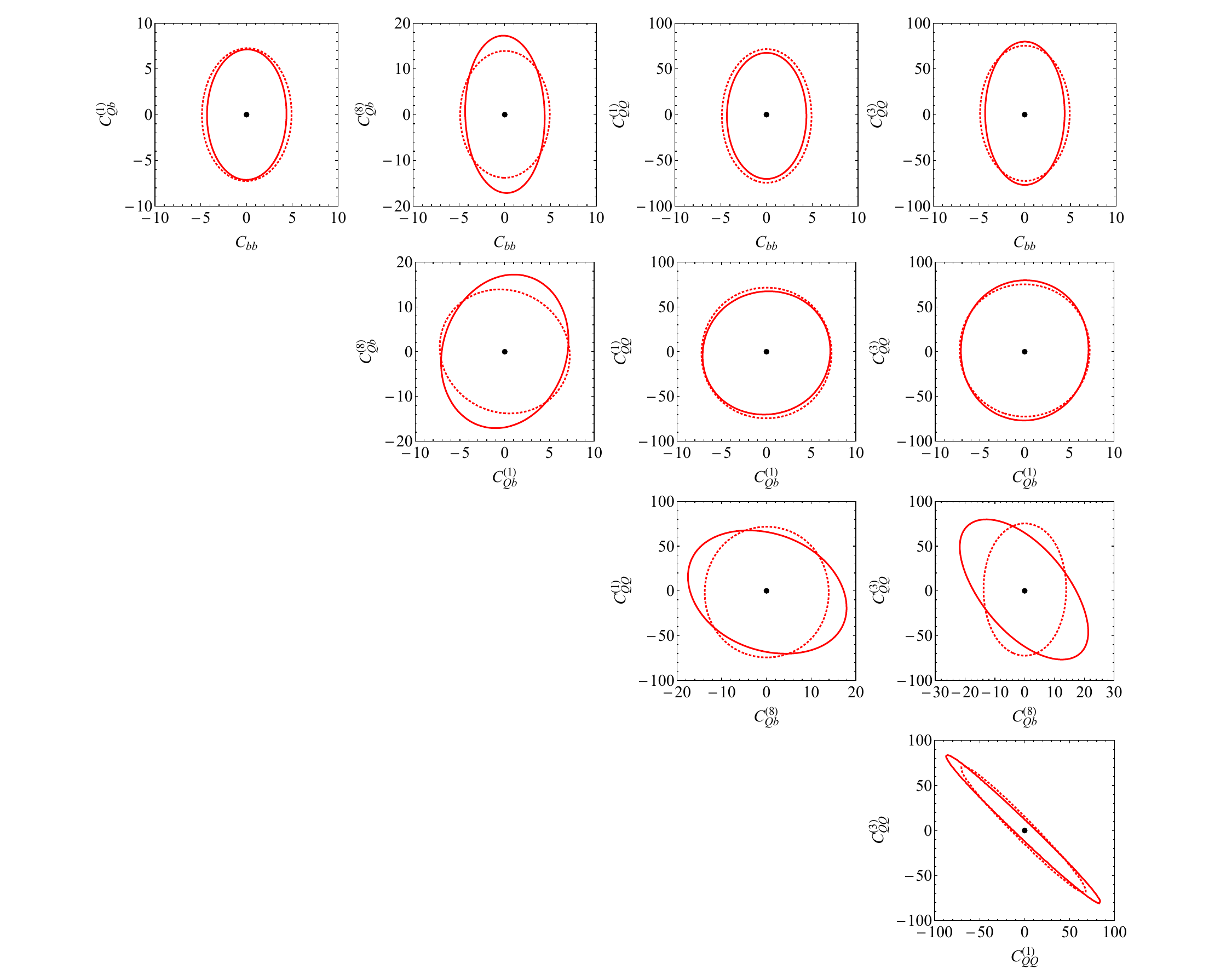} 
\end{center}
\vspace{0mm} 
\caption{$95\%$~CL limits on all possible pairs of third-generation four-quark operators involving four bottom-quark fields. The solid (dotted) red contours show the constraints on the corresponding Wilson~coefficients evaluated at $v = 250 \, {\rm GeV}$ ($\Lambda = 10 \, {\rm TeV}$). The black point in each of the ten panels represents the SM. Additional details can be found in the main text. \label{fig:plots2D}} 
\end{figure}

One-dimensional constraints are extracted from the profiled $\chi^2$ by minimizing it and imposing $\Delta\chi^2 = \chi^2 - \chi^2_{\min} = 3.84$.The resulting limits, in units of $1/{\rm TeV}^2$, are summarized in~Table~\ref{tab:1DCL2}. Note that, for ease of comparison with existing bounds on the Wilson~coefficients of third-generation four-quark operators, all quoted ranges refer to the Wilson~coefficients evaluated at the EW scale $v = 250 \, {\rm GeV}$. The bounds shown are obtained as follows. At~the high scale $\Lambda = 10 \, {\rm TeV}$, each Wilson~coefficient is switched on individually, and the RG~equations are solved in the one-loop~LL~approximation to determine all non-zero Wilson~coefficients at the scale $M_{jj}$. The resulting $\chi^2$ is therefore effectively expressed in terms of Wilson~coefficients evaluated at $M_{jj}$, i.e.,~the relevant scale for the physical process under consideration. The bounds on the high-scale Wilson~coefficients obtained from the~$\chi^2$~fit are finally evolved down to $v$ using, again, the solutions to the RG equations in the one-loop~LL~approximation. The calculation of the logarithmically enhanced corrections fully retains the dependence on the three SM gauge couplings as well as the top-quark Yukawa coupling. The RG running effects correspond to relative shifts of $6\%$, $-6\%$, $24\%$, $13\%$, $8\%$, $-7\%$, $2\%$, $-19\%$, $3\%$, and $-4\%$ in the order of the Wilson~coefficients listed in the table. Overall, Table~\ref{tab:1DCL2} shows that the direct bounds on the Wilson~coefficients of third-generation four-quark operators involving four bottom quarks --- namely, $Q^{(1)}_{QQ}$, $Q^{(3)}_{QQ}$, $Q_{bb}$, $Q^{(1)}_{Qb}$, and $Q^{(8)}_{Qb}$ --- are significantly stronger than the indirect bounds on the Wilson~coefficients of the remaining operators, $Q_{tt}$, $Q^{(1)}_{Qt}$, $Q^{(8)}_{Qt}$, $Q^{(1)}_{tb}$, and~$Q^{(8)}_{tb}$. This demonstrates that the PDF enhancement is not sufficient to overcome the loop suppression of the third-generation four-quark operators, providing only a partial compensation. The most pronounced effect is observed for $Q^{(8)}_{tb}$, which receives~LL~corrections already at the one-loop level, see~(\ref{eq:oneloopLL}).

The $95\%$~CL limits in~Table~\ref{tab:1DCL2} can be compared with similar constraints reported in the literature~\cite{Gauld:2015lmb,Ethier:2021bye,Dawson:2022bxd,Aoude:2022deh,ATLAS:2023ajo,Degrande:2024mbg,Haisch:2024wnw,Haisch:2025vqj,DiNoi:2025uhu}. For $C_{bb}$, $C^{(1)}_{Qb}$, $C^{(8)}_{Qb}$, $C^{(1)}_{tb}$, and $C^{(8)}_{tb}$, our dijet limits provide, to the best of our knowledge, the most stringent constraints to date, whereas the bounds on $C^{(1)}_{QQ}$ and~$C^{(3)}_{QQ}$ are generally weaker. This weakness originates from a flat direction in parameter space, approximately along $C^{(1)}_{QQ} = -C^{(3)}_{QQ}$. This flat direction arises because the linear combination $C^{(1)}_{QQ} + C^{(3)}_{QQ}$ of Wilson~coefficients induces $b \bar b \to t \bar t$ scattering but not $b \bar b \to b \bar b$ scattering at tree level. Once RG effects are included, however, this is no longer the case, and as a result dijet production is partially able to lift the aforementioned degeneracy. Additionally, the limits on $C_{tt}$, $C^{(1)}_{Qt}$, and $C^{(8)}_{Qt}$ are significantly less restrictive than those obtained from EW precision and top-quark measurements~\cite{Ethier:2021bye,Dawson:2022bxd,Aoude:2022deh,ATLAS:2023ajo,Degrande:2024mbg,Haisch:2024wnw,DiNoi:2025uhu}. The~derived $95\%$~CL limits on $C^{(1)}_{QQ}$, $C^{(3)}_{QQ}$, $C_{bb}$, $C^{(1)}_{Qt}$, $C^{(8)}_{Qt}$, $C^{(1)}_{Qb}$, $C^{(8)}_{Qb}$, and $C^{(1)}_{Qt}$ are approximately symmetric, whereas those on~$C_{tt}$ and $C^{(8)}_{tb}$ are noticeably asymmetric, indicating that linear contributions in the former set are subdominant but become significant for the latter. We also note that some of the bounds in Table~\ref{tab:1DCL2} are so weak that the validity of the effective field theory~(EFT) framework used to derive them could be questioned. A detailed quantitative comparison of the one-dimensional constraints on third-generation four-quark operators obtained in this work with those reported in the literature is provided in Appendix~\ref{app:comparison}.

In~Figure~\ref{fig:plots2D}, we present two-dimensional constraints derived from a profiled $\chi^2$~fit, requiring $\Delta \chi^2 = \chi^2 - \chi_{\rm min}^2 = 5.99$. We focus on the Wilson~coefficients $C_{bb}$, $C^{(1)}_{Qb}$, $C^{(8)}_{Qb}$, $C^{(1)}_{QQ}$, and $C^{(3)}_{QQ}$, which, as discussed above, are the five third-generation four-quark operators most strongly constrained by LHC dijet data. The solid~(dotted) red contours indicate the constraints on the Wilson~coefficients evaluated at $v = 250 \, {\rm GeV}$~($\Lambda = 10 \, {\rm TeV}$), and the black point in each panel represents the SM. Two features stand out in the figure: the previously mentioned flat direction, approximately along $C^{(1)}_{QQ} = -C^{(3)}_{QQ}$, and the noticeable impact of one-loop LL corrections from the SMEFT RG flow on the~$95\%$~CL~limits.   The~RG~effects are particularly significant for all the pairs of Wilson~coefficients involving~$C^{(8)}_{Qb}$ due to the strong QCD mixing of $Q^{(8)}_{Qb}$ with all the other relevant third-generation four-quark operators. Note also that in the $C^{(1)}_{QQ}\hspace{0.25mm}$--$\hspace{0.25mm}C^{(3)}_{QQ}$ plane closed 95\%~CL regions emerge only after the inclusion of radiative corrections. These findings demonstrate that properly incorporating RG-induced effects is essential when relating LHC and low-energy SMEFT constraints, as~RG~running can significantly modify the limits on Wilson~coefficients at different energy scales and help resolve flat directions in the multi-dimensional space of Wilson~coefficients. As in the one-dimensional case, we have explicitly verified that marginalization produces very similar, though slightly stronger, constraints in the planes of Wilson~coefficient pairs shown in the~ten panels of~Figure~\ref{fig:plots2D}. 

\section{Conclusions}
\label{sec:conclusions}

In this article, we have investigated the sensitivity of LHC dijet measurements to the ten third-generation four-quark operators in the SMEFT listed in~(\ref{eq:4Foperators}). At tree level, only the five operators involving four bottom quarks, $Q^{(1)}_{QQ}$, $Q^{(3)}_{QQ}$, $Q_{bb}$, $Q^{(1)}_{Qb}$, and $Q^{(8)}_{Qb}$, contribute directly to dijet production. Nevertheless, RG effects that generate logarithmically enhanced contributions allow all ten operators to be probed indirectly when one- and two-loop corrections are included. These contributions stem from QCD and QED penguin diagrams, which mediate flavor-changing effects among the four-quark operators.

Using the normalized dijet angular distributions measured by CMS in~\cite{CMS:2018ucw}, we have derived one-dimensional $95\%$ CL limits on the Wilson~coefficients of the full set of third-generation four-quark operators. The bounds for operators involving four bottom quarks are relatively strong, often improving upon or comparable to existing limits reported in the literature~\cite{Gauld:2015lmb,Ethier:2021bye,Dawson:2022bxd,Aoude:2022deh,ATLAS:2023ajo,Degrande:2024mbg,Haisch:2024wnw,Haisch:2025vqj,DiNoi:2025uhu}. In contrast, the constraints on the remaining third-generation four-quark operators, which involve only top quarks or both bottom and top quarks, remain comparatively weak and are more effectively probed via EW precision measurements and top-quark observables at the LHC~\cite{Ethier:2021bye,Dawson:2022bxd,Aoude:2022deh,ATLAS:2023ajo,Degrande:2024mbg,Haisch:2024wnw,DiNoi:2025uhu}. Our findings demonstrate that the enhancement from PDFs, arising from the RG-induced flavor transition from third-generation to first- or second-generation fields, is insufficient to overcome the loop suppression of these operators. For the Wilson~coefficients $C^{(1)}_{QQ}$, $C^{(3)}_{QQ}$, $C_{bb}$, $C^{(1)}_{Qb}$, and $C^{(8)}_{Qb}$, we have also derived two-dimensional constraints for all possible pairs. The analysis shows that RG effects between the initial scale $\Lambda$ and the EW scale $v$ are especially significant for the Wilson-coefficient pairs involving $C^{(8)}_{Qb}$. This is driven by the strong QCD mixing of $Q^{(8)}_{Qb}$ with all other relevant third-generation four-quark operators. This result underscores the importance of including RG-induced contributions when interpreting high-energy LHC data in terms of SMEFT operators, as they can significantly affect the derived constraints at the~EW scale.

Due to the high values of the jet-jet invariant mass $M_{jj}$ probed in LHC dijet analyses, the constraints on the Wilson~coefficients of the third-generation four-quark operators derived in this work are generally dominated by the quadratic BSM contributions, with interference between the SM and BSM terms in most cases playing only a subleading role. As a consequence, the derived limits are sensitive to dimension-eight deformations of the~SMEFT. This issue, which also affects similar constraints from $t \bar t$, $t \bar t t \bar t$, and $t \bar t b \bar b$ production at the LHC~\cite{Ethier:2021bye,Aoude:2022deh,ATLAS:2023ajo,Degrande:2024mbg,DiNoi:2025uhu,ATLAS:2018fwl,CMS:2019rvj,CMS:2019eih,ATLAS:2020hpj,Banelli:2020iau,ATLAS:2021kqb,CMS:2023zdh,CMS-PAS-TOP-24-008}, but not the limits from Higgs or EW precision measurements~\cite{Dawson:2022bxd,Haisch:2024wnw,Haisch:2025vqj,DiNoi:2025uhu}, should be kept in mind when interpreting the derived bounds in terms of explicit UV~complete~models. Moreover, the relative weakness of some of the bounds shown in~Table~\ref{tab:1DCL2}~and~Figure~\ref{fig:plots2D} raises concerns regarding the validity of the EFT approach employed in this analysis. Ultimately, this issue can only be addressed through the study of concrete BSM scenarios with dominant couplings to third-generation SM fermions, which is, however, beyond the scope of the present work.

\acknowledgments{UH gratefully acknowledges Ben Stefanek, fellow enthusiast of SMEFT double logarithms, for enjoyable discussions. The analytic calculations of the squared matrix elements relevant to this study were performed using {\tt FeynRules}~\cite{Alloul:2013bka}, {\tt FeynArts}~\cite{Hahn:2000kx}, and {\tt FormCalc}~\cite{Hahn:2016ebn}, while the beta functions implemented in {\tt DsixTools~2.0}~\cite{Fuentes-Martin:2020zaz} were used to derive the LL-enhanced effects up to the two-loop order.}

\begin{appendix}

\section{Compendium of constraints} 
\label{app:comparison}

In this appendix, we compile existing constraints from the literature on the third-generation four-quark operators introduced in~(\ref{eq:4Foperators}). Since most of the literature does not use the Warsaw basis but rather the {\tt dim6top} basis~\cite{Aguilar-Saavedra:2018ksv}, we translate the bounds given in Table~\ref{tab:1DCL2}, which are expressed in the former basis, into the latter. While this translation is generally non-trivial, as it involves a Fierz-evanescent operator~\cite{Degrande:2024mbg,Haisch:2024wnw}, for the constraints derived in this article it suffices to perform the following linear transformations:
\begin{gather}
c_{QQ}^{(1)} = 2 \hspace{0.125mm} C_{QQ}^{(1)} - \frac{2}{3} \hspace{0.25mm} C_{QQ}^{(3)} \,, \qquad c_{QQ}^{(8)} = 8 \hspace{0.125mm} C_{QQ}^{(3)} \,, \nonumber \\[2mm] 
c_{tt} = C_{tt} \,, \qquad c_{bb} = C_{bb} \,, \qquad c_{Qt}^{(1)} = C_{Qt}^{(1)} \,, \qquad c_{Qt}^{(8)} = C_{Qt}^{(8)} \,, \label{eq:change} \\[2mm]
c_{Qb}^{(1)} = C_{Qb}^{(1)} \,, \qquad c_{Qb}^{(8)} = C_{Qb}^{(8)} \,, \qquad c_{tb}^{(1)} = C_{tb}^{(1)} \,, \qquad c_{tb}^{(8)} = C_{tb}^{(8)} \,. \nonumber
\end{gather}
Here, the Wilson coefficients on the right-hand sides correspond to the Warsaw basis, while those on the left-hand sides correspond to the {\tt dim6top} basis. We emphasize that the same change of basis in~(\ref{eq:change}) can be applied to all constraints in the literature that arise from tree-level processes, such as $t \bar t t \bar t$ production. However, it generally misses finite contributions proportional to $c_{QQ}^{(8)}$ when one-loop corrections to $pp \to t \bar t$ or $Z \to b \bar b$ are taken into account~\cite{Degrande:2024mbg,Haisch:2024wnw}. In global Wilson coefficient analyses such as~\cite{DiNoi:2025uhu}, which combine tree- and loop-level constraints, it is therefore crucial to use a consistent definition of operators across all calculations to achieve well-defined fit results. 

{
\renewcommand{\arraystretch}{1.25}
\begin{table}[t!]
\scriptsize
\centering
\begin{tabular}{|c|c|c|c|c|c|c|c|}
\hline
& this work & \cite{Haisch:2025vqj} & \cite{Ethier:2021bye} & \cite{Dawson:2022bxd} & \cite{ATLAS:2023ajo} & \cite{Degrande:2024mbg} & \cite{DiNoi:2025uhu} \\ \hline \hline 
$c^{(1)}_{QQ}$ & $[-67,63]$ & -- & $[-3.0,3.7]$ & $[-1.6,2.7]$ & $[-4.0,4.5]$ & \cellcolor{gray!25}$[-2.5,3.9]$ & $[-0.97,5.9]$ \\ \hline  
$c^{(8)}_{QQ}$ & $[-59,63]$ & -- &  $[-11,8.2]$ & $[-15,25]$ & -- & \cellcolor{gray!25}$[-9.9,5.5]$ & $[-19,5.8]$ \\ \hline  
$c_{tt}$ & $[-850,330]$ & -- & $[-0.79,0.71]$ & -- & $[-1.9,2.1]$ & \cellcolor{gray!25}$[-0.71,0.80]$ & $[ -1.5,1.5]$ \\ \hline  
$c_{bb}$ & \cellcolor{gray!25}$[-4.7,4.7]$ & -- & -- &  $[-360, 95]$ & -- & -- & -- \\ \hline  
$c^{(1)}_{Qt}$ & $[-290,300]$ & $[-4.2,23]$ & $[-1.4,1.3]$ & $[-2.2,1.3]$ & $[-3.8,3.4]$ & \cellcolor{gray!25}$[-1.4,1.2]$ & $[-1.4,2.7]$ \\ \hline  
$c^{(8)}_{Qt}$ & $[-490,500]$ & $[-17,90]$ &  $[-3.0,2.2]$ & -- & $[-6.9,7.6]$ & \cellcolor{gray!25}$[-2.5,2.6]$ & $[-4.1,6.2]$ \\ \hline  
$c^{(1)}_{Qb}$ & \cellcolor{gray!25}$[-6.2,6.2]$ & $[-1030,190]$ & -- & $[-21,4.9]$ & -- & -- & -- \\ \hline  
$c^{(8)}_{Qb}$ & \cellcolor{gray!25}$[-9.3,9.4]$ & $[-880,160]$ & -- & -- & -- & -- & -- \\ \hline  
$c^{(1)}_{tb}$ & $[-310,310]$ & -- & -- & \cellcolor{gray!25}$[-4.1,16]$ & -- & -- & -- \\ \hline  
$c^{(8)}_{tb}$ & \cellcolor{gray!25}$[-130, 170]$ & -- & -- & -- & -- & -- & -- \\ \hline 
\end{tabular}
\vspace{4mm} 
\caption{Compendium of $95\%$ CL limits on the Wilson coefficients of third-generation four-quark operators in the {\tt dim6top} basis, given in units of $1/{\rm TeV}^2$. Entries marked with ``--'' indicate that no constraints are reported in the corresponding work. Cells corresponding to the tightest nominal bounds for each Wilson coefficient are highlighted in gray. See the main text for further details. \label{tab:comparison}}
\end{table}
}

Table~\ref{tab:comparison} summarizes the $95\%$ CL limits on the Wilson coefficients of third-generation four-quark operators in the {\tt dim6top} basis, with values given in units of $1/{\rm TeV}^2$. To allow a meaningful comparison among the different results, it is useful to briefly describe how the bounds are obtained. In the article~\cite{Haisch:2025vqj}, Higgs production in $gg \to h$, which becomes sensitive to certain third-generation four-quark operators at the two-loop level~\cite{DiNoi:2023ygk,Haisch:2025vqj}, is used to derive individual bounds on the relevant Wilson coefficients. In contrast,~\cite{Ethier:2021bye} performs a global fit to LHC data on $t\bar t$, $t\bar t t\bar t$, and $t\bar t b\bar b$ production, with the reported constraints corresponding to marginalized limits that include quadratic terms in the Wilson coefficients. NLO corrections are included for $t\bar t$ production, while $t\bar t t\bar t$ and $t\bar t b\bar b$ production are evaluated at LO. One-loop corrections to $Z \to b\bar b$ with single insertions of third-generation four-quark operators are computed in~\cite{Dawson:2022bxd} and used to derive individual constraints on the corresponding Wilson coefficients. The ATLAS analysis~\cite{ATLAS:2023ajo} studies $t\bar t t\bar t$ production at tree level and extracts individual bounds including quadratic contributions in the Wilson coefficients. A global analysis of $t\bar t$ and $t\bar t t\bar t$ data is presented in~\cite{Degrande:2024mbg}, where the reported limits are marginalized and include quadratic terms in the Wilson coefficients, based on NLO~(LO)~QCD predictions for $t\bar t$ ($t\bar t t\bar t$) production. The~comprehensive global fit of \cite{DiNoi:2025uhu} combines NLO predictions for $gg \to h$, $t\bar t$, and~$t\bar t h$ production with LO calculations for $t\bar t t\bar t$ and $t\bar t b\bar b$, and incorporates one-loop corrections to~$Z \to b\bar b$ as well as two-loop contributions to the Peskin–Takeuchi parameters, both taken from \cite{Haisch:2024wnw}. The resulting limits are marginalized and include quadratic terms in the Wilson~coefficients.

The cells with the tightest nominal bounds for each Wilson coefficient are highlighted in gray in Table~\ref{tab:comparison}. For $c^{(1)}_{QQ}$, $c^{(8)}_{QQ}$, $c_{tt}$, $c^{(1)}_{Qt}$, and $c^{(8)}_{Qt}$, the tightest bounds are those reported in~\cite{Degrande:2024mbg}, obtained from detailed analyses of $t\bar t$ and $t\bar t t\bar t$ production. However, the quoted marginalized limits depend non-negligibly on the inclusion of quadratic terms in the Wilson coefficients, with the quadratic analysis yielding stronger constraints due to sensitivity to the high-energy tails of the distributions, an undesirable feature from the perspective of EFT validity. In this context, we note that the limits from~\cite{DiNoi:2025uhu}, though nominally weaker than those given in~\cite{Degrande:2024mbg}, are more robust due to the inclusion of one- and two-loop corrections to EWPOs~\cite{Haisch:2024wnw}, which tighten linear-fit bounds and lift degeneracies in the quadratic~fit. The importance of EWPOs is further illustrated for $c_{tb}^{1}$, where the one-loop analysis of~$Z \to b\bar{b}$ in~\cite{Dawson:2022bxd} currently provides the strongest constraint. For~the remaining coefficients, $c_{bb}$, $c^{(1)}_{Qb}$, $c^{(8)}_{Qb}$, and $c^{(8)}_{tb}$, Table~\ref{tab:comparison} shows that the dijet analysis presented here provides the most stringent limits. Taken together, these findings show that robust constraints on the Wilson coefficients of the complete set of third-generation four-quark operators~(\ref{eq:4Foperators}) requires a global fit combining heavy-quark production, EWPOs, and dijet~data, exploiting the complementarity and synergy of these different observables.

\end{appendix}


\begin{thebibliography}{73}%
\makeatletter
\providecommand \@ifxundefined [1]{%
 \@ifx{#1\undefined}
}%
\providecommand \@ifnum [1]{%
 \ifnum #1\expandafter \@firstoftwo
 \else \expandafter \@secondoftwo
 \fi
}%
\providecommand \@ifx [1]{%
 \ifx #1\expandafter \@firstoftwo
 \else \expandafter \@secondoftwo
 \fi
}%
\providecommand \natexlab [1]{#1}%
\providecommand \enquote  [1]{``#1''}%
\providecommand \bibnamefont  [1]{#1}%
\providecommand \bibfnamefont [1]{#1}%
\providecommand \citenamefont [1]{#1}%
\providecommand \href@noop [0]{\@secondoftwo}%
\providecommand \href [0]{\begingroup \@sanitize@url \@href}%
\providecommand \@href[1]{\@@startlink{#1}\@@href}%
\providecommand \@@href[1]{\endgroup#1\@@endlink}%
\providecommand \@sanitize@url [0]{\catcode `\\12\catcode `\$12\catcode
  `\&12\catcode `\#12\catcode `\^12\catcode `\_12\catcode `\%12\relax}%
\providecommand \@@startlink[1]{}%
\providecommand \@@endlink[0]{}%
\providecommand \url  [0]{\begingroup\@sanitize@url \@url }%
\providecommand \@url [1]{\endgroup\@href {#1}{\urlprefix }}%
\providecommand \urlprefix  [0]{URL }%
\providecommand \Eprint [0]{\href }%
\providecommand \doibase [0]{http://dx.doi.org/}%
\providecommand \selectlanguage [0]{\@gobble}%
\providecommand \bibinfo  [0]{\@secondoftwo}%
\providecommand \bibfield  [0]{\@secondoftwo}%
\providecommand \translation [1]{[#1]}%
\providecommand \BibitemOpen [0]{}%
\providecommand \bibitemStop [0]{}%
\providecommand \bibitemNoStop [0]{.\EOS\space}%
\providecommand \EOS [0]{\spacefactor3000\relax}%
\providecommand\BibitemShut  [1]{\csname bibitem#1\endcsname}%
\let\auto@bib@innerbib\@empty
\bibitem [{\citenamefont {Buchm{\"u}ller}\ and\ \citenamefont
  {Wyler}(1986)}]{Buchmuller:1985jz}%
  \BibitemOpen
  \bibfield  {author} {\bibinfo {author} {\bibfnamefont {W.}~\bibnamefont
  {Buchm{\"u}ller}}\ and\ \bibinfo {author} {\bibfnamefont {D.}~\bibnamefont
  {Wyler}},\ }\href {\doibase 10.1016/0550-3213(86)90262-2} {\bibfield
  {journal} {\bibinfo  {journal} {Nucl. Phys. B}\ }\textbf {\bibinfo {volume}
  {268}},\ \bibinfo {pages} {621} (\bibinfo {year} {1986})}\BibitemShut
  {NoStop}%
\bibitem [{\citenamefont {Grzadkowski}\ \emph {et~al.}(2010)\citenamefont
  {Grzadkowski}, \citenamefont {Iskrzynski}, \citenamefont {Misiak},\ and\
  \citenamefont {Rosiek}}]{Grzadkowski:2010es}%
  \BibitemOpen
  \bibfield  {author} {\bibinfo {author} {\bibfnamefont {B.}~\bibnamefont
  {Grzadkowski}}, \bibinfo {author} {\bibfnamefont {M.}~\bibnamefont
  {Iskrzynski}}, \bibinfo {author} {\bibfnamefont {M.}~\bibnamefont {Misiak}},
  \ and\ \bibinfo {author} {\bibfnamefont {J.}~\bibnamefont {Rosiek}},\ }\href
  {\doibase 10.1007/JHEP10(2010)085} {\bibfield  {journal} {\bibinfo  {journal}
  {JHEP}\ }\textbf {\bibinfo {volume} {10}},\ \bibinfo {pages} {085} (\bibinfo
  {year} {2010})},\ \Eprint {http://arxiv.org/abs/1008.4884} {arXiv:1008.4884
  [hep-ph]}\BibitemShut {NoStop}%
\bibitem [{\citenamefont {Brivio}\ and\ \citenamefont
  {Trott}(2019)}]{Brivio:2017vri}%
  \BibitemOpen
  \bibfield  {author} {\bibinfo {author} {\bibfnamefont {I.}~\bibnamefont
  {Brivio}}\ and\ \bibinfo {author} {\bibfnamefont {M.}~\bibnamefont {Trott}},\
  }\href {\doibase 10.1016/j.physrep.2018.11.002} {\bibfield  {journal}
  {\bibinfo  {journal} {Phys. Rept.}\ }\textbf {\bibinfo {volume} {793}},\
  \bibinfo {pages} {1} (\bibinfo {year} {2019})},\ \Eprint
  {http://arxiv.org/abs/1706.08945} {arXiv:1706.08945 [hep-ph]}\BibitemShut
  {NoStop}%
\bibitem [{\citenamefont {Isidori}\ \emph {et~al.}(2024)\citenamefont
  {Isidori}, \citenamefont {Wilsch},\ and\ \citenamefont
  {Wyler}}]{Isidori:2023pyp}%
  \BibitemOpen
  \bibfield  {author} {\bibinfo {author} {\bibfnamefont {G.}~\bibnamefont
  {Isidori}}, \bibinfo {author} {\bibfnamefont {F.}~\bibnamefont {Wilsch}}, \
  and\ \bibinfo {author} {\bibfnamefont {D.}~\bibnamefont {Wyler}},\ }\href
  {\doibase 10.1103/RevModPhys.96.015006} {\bibfield  {journal} {\bibinfo
  {journal} {Rev. Mod. Phys.}\ }\textbf {\bibinfo {volume} {96}},\ \bibinfo
  {pages} {015006} (\bibinfo {year} {2024})},\ \Eprint
  {http://arxiv.org/abs/2303.16922} {arXiv:2303.16922 [hep-ph]}\BibitemShut
  {NoStop}%
\bibitem [{\citenamefont {Jenkins}\ \emph {et~al.}(2013)\citenamefont
  {Jenkins}, \citenamefont {Manohar},\ and\ \citenamefont
  {Trott}}]{Jenkins:2013zja}%
  \BibitemOpen
  \bibfield  {author} {\bibinfo {author} {\bibfnamefont {E.~E.}\ \bibnamefont
  {Jenkins}}, \bibinfo {author} {\bibfnamefont {A.~V.}\ \bibnamefont
  {Manohar}}, \ and\ \bibinfo {author} {\bibfnamefont {M.}~\bibnamefont
  {Trott}},\ }\href {\doibase 10.1007/JHEP10(2013)087} {\bibfield  {journal}
  {\bibinfo  {journal} {JHEP}\ }\textbf {\bibinfo {volume} {10}},\ \bibinfo
  {pages} {087} (\bibinfo {year} {2013})},\ \Eprint
  {http://arxiv.org/abs/1308.2627} {arXiv:1308.2627 [hep-ph]}\BibitemShut
  {NoStop}%
\bibitem [{\citenamefont {Jenkins}\ \emph {et~al.}(2014)\citenamefont
  {Jenkins}, \citenamefont {Manohar},\ and\ \citenamefont
  {Trott}}]{Jenkins:2013wua}%
  \BibitemOpen
  \bibfield  {author} {\bibinfo {author} {\bibfnamefont {E.~E.}\ \bibnamefont
  {Jenkins}}, \bibinfo {author} {\bibfnamefont {A.~V.}\ \bibnamefont
  {Manohar}}, \ and\ \bibinfo {author} {\bibfnamefont {M.}~\bibnamefont
  {Trott}},\ }\href {\doibase 10.1007/JHEP01(2014)035} {\bibfield  {journal}
  {\bibinfo  {journal} {JHEP}\ }\textbf {\bibinfo {volume} {01}},\ \bibinfo
  {pages} {035} (\bibinfo {year} {2014})},\ \Eprint
  {http://arxiv.org/abs/1310.4838} {arXiv:1310.4838 [hep-ph]}\BibitemShut
  {NoStop}%
\bibitem [{\citenamefont {Alonso}\ \emph {et~al.}(2014)\citenamefont {Alonso},
  \citenamefont {Jenkins}, \citenamefont {Manohar},\ and\ \citenamefont
  {Trott}}]{Alonso:2013hga}%
  \BibitemOpen
  \bibfield  {author} {\bibinfo {author} {\bibfnamefont {R.}~\bibnamefont
  {Alonso}}, \bibinfo {author} {\bibfnamefont {E.~E.}\ \bibnamefont {Jenkins}},
  \bibinfo {author} {\bibfnamefont {A.~V.}\ \bibnamefont {Manohar}}, \ and\
  \bibinfo {author} {\bibfnamefont {M.}~\bibnamefont {Trott}},\ }\href
  {\doibase 10.1007/JHEP04(2014)159} {\bibfield  {journal} {\bibinfo  {journal}
  {JHEP}\ }\textbf {\bibinfo {volume} {04}},\ \bibinfo {pages} {159} (\bibinfo
  {year} {2014})},\ \Eprint {http://arxiv.org/abs/1312.2014} {arXiv:1312.2014
  [hep-ph]}\BibitemShut {NoStop}%
\bibitem [{\citenamefont {Gorbahn}\ and\ \citenamefont
  {Haisch}(2016)}]{Gorbahn:2016uoy}%
  \BibitemOpen
  \bibfield  {author} {\bibinfo {author} {\bibfnamefont {M.}~\bibnamefont
  {Gorbahn}}\ and\ \bibinfo {author} {\bibfnamefont {U.}~\bibnamefont
  {Haisch}},\ }\href {\doibase 10.1007/JHEP10(2016)094} {\bibfield  {journal}
  {\bibinfo  {journal} {JHEP}\ }\textbf {\bibinfo {volume} {10}},\ \bibinfo
  {pages} {094} (\bibinfo {year} {2016})},\ \Eprint
  {http://arxiv.org/abs/1607.03773} {arXiv:1607.03773 [hep-ph]}\BibitemShut
  {NoStop}%
\bibitem [{\citenamefont {Bern}\ \emph {et~al.}(2020)\citenamefont {Bern},
  \citenamefont {Parra-Martinez},\ and\ \citenamefont {Sawyer}}]{Bern:2020ikv}%
  \BibitemOpen
  \bibfield  {author} {\bibinfo {author} {\bibfnamefont {Z.}~\bibnamefont
  {Bern}}, \bibinfo {author} {\bibfnamefont {J.}~\bibnamefont
  {Parra-Martinez}}, \ and\ \bibinfo {author} {\bibfnamefont {E.}~\bibnamefont
  {Sawyer}},\ }\href {\doibase 10.1007/JHEP10(2020)211} {\bibfield  {journal}
  {\bibinfo  {journal} {JHEP}\ }\textbf {\bibinfo {volume} {10}},\ \bibinfo
  {pages} {211} (\bibinfo {year} {2020})},\ \Eprint
  {http://arxiv.org/abs/2005.12917} {arXiv:2005.12917 [hep-ph]}\BibitemShut
  {NoStop}%
\bibitem [{\citenamefont {Jin}\ \emph {et~al.}(2021)\citenamefont {Jin},
  \citenamefont {Ren},\ and\ \citenamefont {Yang}}]{Jin:2020pwh}%
  \BibitemOpen
  \bibfield  {author} {\bibinfo {author} {\bibfnamefont {Q.}~\bibnamefont
  {Jin}}, \bibinfo {author} {\bibfnamefont {K.}~\bibnamefont {Ren}}, \ and\
  \bibinfo {author} {\bibfnamefont {G.}~\bibnamefont {Yang}},\ }\href {\doibase
  10.1007/JHEP04(2021)180} {\bibfield  {journal} {\bibinfo  {journal} {JHEP}\
  }\textbf {\bibinfo {volume} {04}},\ \bibinfo {pages} {180} (\bibinfo {year}
  {2021})},\ \Eprint {http://arxiv.org/abs/2011.02494} {arXiv:2011.02494
  [hep-ph]}\BibitemShut {NoStop}%
\bibitem [{\citenamefont {Haisch}\ \emph {et~al.}(2022)\citenamefont {Haisch},
  \citenamefont {Scott}, \citenamefont {Wiesemann}, \citenamefont
  {Zanderighi},\ and\ \citenamefont {Zanoli}}]{Haisch:2022nwz}%
  \BibitemOpen
  \bibfield  {author} {\bibinfo {author} {\bibfnamefont {U.}~\bibnamefont
  {Haisch}}, \bibinfo {author} {\bibfnamefont {D.~J.}\ \bibnamefont {Scott}},
  \bibinfo {author} {\bibfnamefont {M.}~\bibnamefont {Wiesemann}}, \bibinfo
  {author} {\bibfnamefont {G.}~\bibnamefont {Zanderighi}}, \ and\ \bibinfo
  {author} {\bibfnamefont {S.}~\bibnamefont {Zanoli}},\ }\href {\doibase
  10.1007/JHEP07(2022)054} {\bibfield  {journal} {\bibinfo  {journal} {JHEP}\
  }\textbf {\bibinfo {volume} {07}},\ \bibinfo {pages} {054} (\bibinfo {year}
  {2022})},\ \Eprint {http://arxiv.org/abs/2204.00663} {arXiv:2204.00663
  [hep-ph]}\BibitemShut {NoStop}%
\bibitem [{\citenamefont {Di~Noi}\ \emph {et~al.}(2024)\citenamefont {Di~Noi},
  \citenamefont {Gr\"ober}, \citenamefont {Heinrich}, \citenamefont {Lang},\
  and\ \citenamefont {Vitti}}]{DiNoi:2023ygk}%
  \BibitemOpen
  \bibfield  {author} {\bibinfo {author} {\bibfnamefont {S.}~\bibnamefont
  {Di~Noi}}, \bibinfo {author} {\bibfnamefont {R.}~\bibnamefont {Gr\"ober}},
  \bibinfo {author} {\bibfnamefont {G.}~\bibnamefont {Heinrich}}, \bibinfo
  {author} {\bibfnamefont {J.}~\bibnamefont {Lang}}, \ and\ \bibinfo {author}
  {\bibfnamefont {M.}~\bibnamefont {Vitti}},\ }\href {\doibase
  10.1103/PhysRevD.109.095024} {\bibfield  {journal} {\bibinfo  {journal}
  {Phys. Rev. D}\ }\textbf {\bibinfo {volume} {109}},\ \bibinfo {pages}
  {095024} (\bibinfo {year} {2024})},\ \Eprint
  {http://arxiv.org/abs/2310.18221} {arXiv:2310.18221 [hep-ph]}\BibitemShut
  {NoStop}%
\bibitem [{\citenamefont {Jenkins}\ \emph {et~al.}(2024)\citenamefont
  {Jenkins}, \citenamefont {Manohar}, \citenamefont {Naterop},\ and\
  \citenamefont {Pag\`es}}]{Jenkins:2023bls}%
  \BibitemOpen
  \bibfield  {author} {\bibinfo {author} {\bibfnamefont {E.~E.}\ \bibnamefont
  {Jenkins}}, \bibinfo {author} {\bibfnamefont {A.~V.}\ \bibnamefont
  {Manohar}}, \bibinfo {author} {\bibfnamefont {L.}~\bibnamefont {Naterop}}, \
  and\ \bibinfo {author} {\bibfnamefont {J.}~\bibnamefont {Pag\`es}},\ }\href
  {\doibase 10.1007/JHEP02(2024)131} {\bibfield  {journal} {\bibinfo  {journal}
  {JHEP}\ }\textbf {\bibinfo {volume} {02}},\ \bibinfo {pages} {131} (\bibinfo
  {year} {2024})},\ \Eprint {http://arxiv.org/abs/2310.19883} {arXiv:2310.19883
  [hep-ph]}\BibitemShut {NoStop}%
\bibitem [{\citenamefont {Di~Noi}\ \emph
  {et~al.}(2025{\natexlab{a}})\citenamefont {Di~Noi}, \citenamefont
  {Gr\"ober},\ and\ \citenamefont {Mandal}}]{DiNoi:2024ajj}%
  \BibitemOpen
  \bibfield  {author} {\bibinfo {author} {\bibfnamefont {S.}~\bibnamefont
  {Di~Noi}}, \bibinfo {author} {\bibfnamefont {R.}~\bibnamefont {Gr\"ober}}, \
  and\ \bibinfo {author} {\bibfnamefont {M.~K.}\ \bibnamefont {Mandal}},\
  }\href {\doibase 10.1007/JHEP12(2024)220} {\bibfield  {journal} {\bibinfo
  {journal} {JHEP}\ }\textbf {\bibinfo {volume} {12}},\ \bibinfo {pages} {220}
  (\bibinfo {year} {2025}{\natexlab{a}})},\ \Eprint
  {http://arxiv.org/abs/2408.03252} {arXiv:2408.03252 [hep-ph]}\BibitemShut
  {NoStop}%
\bibitem [{\citenamefont {Born}\ \emph {et~al.}(2025)\citenamefont {Born},
  \citenamefont {Fuentes-Mart{\'\i}n}, \citenamefont {Kvedarait{\.{e}}},\ and\
  \citenamefont {Thomsen}}]{Born:2024mgz}%
  \BibitemOpen
  \bibfield  {author} {\bibinfo {author} {\bibfnamefont {L.}~\bibnamefont
  {Born}}, \bibinfo {author} {\bibfnamefont {J.}~\bibnamefont
  {Fuentes-Mart{\'\i}n}}, \bibinfo {author} {\bibfnamefont {S.}~\bibnamefont
  {Kvedarait{\.{e}}}}, \ and\ \bibinfo {author} {\bibfnamefont {A.~E.}\
  \bibnamefont {Thomsen}},\ }\href {\doibase 10.1007/JHEP05(2025)121}
  {\bibfield  {journal} {\bibinfo  {journal} {JHEP}\ }\textbf {\bibinfo
  {volume} {05}},\ \bibinfo {pages} {121} (\bibinfo {year} {2025})},\ \Eprint
  {http://arxiv.org/abs/2410.07320} {arXiv:2410.07320 [hep-ph]}\BibitemShut
  {NoStop}%
\bibitem [{\citenamefont {Naterop}\ and\ \citenamefont
  {Stoffer}(2025{\natexlab{a}})}]{Naterop:2024cfx}%
  \BibitemOpen
  \bibfield  {author} {\bibinfo {author} {\bibfnamefont {L.}~\bibnamefont
  {Naterop}}\ and\ \bibinfo {author} {\bibfnamefont {P.}~\bibnamefont
  {Stoffer}},\ }\href {\doibase 10.1007/JHEP06(2025)007} {\bibfield  {journal}
  {\bibinfo  {journal} {JHEP}\ }\textbf {\bibinfo {volume} {06}},\ \bibinfo
  {pages} {007} (\bibinfo {year} {2025}{\natexlab{a}})},\ \Eprint
  {http://arxiv.org/abs/2412.13251} {arXiv:2412.13251 [hep-ph]}\BibitemShut
  {NoStop}%
\bibitem [{\citenamefont {Duhr}\ \emph
  {et~al.}(2025{\natexlab{a}})\citenamefont {Duhr}, \citenamefont {Vasquez},
  \citenamefont {Ventura},\ and\ \citenamefont {Vryonidou}}]{Duhr:2025zqw}%
  \BibitemOpen
  \bibfield  {author} {\bibinfo {author} {\bibfnamefont {C.}~\bibnamefont
  {Duhr}}, \bibinfo {author} {\bibfnamefont {A.}~\bibnamefont {Vasquez}},
  \bibinfo {author} {\bibfnamefont {G.}~\bibnamefont {Ventura}}, \ and\
  \bibinfo {author} {\bibfnamefont {E.}~\bibnamefont {Vryonidou}},\ }\href
  {\doibase 10.1007/JHEP07(2025)160} {\bibfield  {journal} {\bibinfo  {journal}
  {JHEP}\ }\textbf {\bibinfo {volume} {07}},\ \bibinfo {pages} {160} (\bibinfo
  {year} {2025}{\natexlab{a}})},\ \Eprint {http://arxiv.org/abs/2503.01954}
  {arXiv:2503.01954 [hep-ph]}\BibitemShut {NoStop}%
\bibitem [{\citenamefont {Haisch}(2025)}]{Haisch:2025lvd}%
  \BibitemOpen
  \bibfield  {author} {\bibinfo {author} {\bibfnamefont {U.}~\bibnamefont
  {Haisch}},\ }\href {\doibase 10.1007/JHEP06(2025)004} {\bibfield  {journal}
  {\bibinfo  {journal} {JHEP}\ }\textbf {\bibinfo {volume} {06}},\ \bibinfo
  {pages} {004} (\bibinfo {year} {2025})},\ \Eprint
  {http://arxiv.org/abs/2503.06249} {arXiv:2503.06249 [hep-ph]}\BibitemShut
  {NoStop}%
\bibitem [{\citenamefont {Zhang}(2025)}]{Zhang:2025ywe}%
  \BibitemOpen
  \bibfield  {author} {\bibinfo {author} {\bibfnamefont {D.}~\bibnamefont
  {Zhang}},\ }\href {\doibase 10.1007/JHEP06(2025)106} {\bibfield  {journal}
  {\bibinfo  {journal} {JHEP}\ }\textbf {\bibinfo {volume} {06}},\ \bibinfo
  {pages} {106} (\bibinfo {year} {2025})},\ \Eprint
  {http://arxiv.org/abs/2504.00792} {arXiv:2504.00792 [hep-ph]}\BibitemShut
  {NoStop}%
\bibitem [{\citenamefont {Assi}\ \emph {et~al.}(2025)\citenamefont {Assi},
  \citenamefont {Helset}, \citenamefont {Pag{\`e}s},\ and\ \citenamefont
  {Shen}}]{Assi:2025fsm}%
  \BibitemOpen
  \bibfield  {author} {\bibinfo {author} {\bibfnamefont {B.}~\bibnamefont
  {Assi}}, \bibinfo {author} {\bibfnamefont {A.}~\bibnamefont {Helset}},
  \bibinfo {author} {\bibfnamefont {J.}~\bibnamefont {Pag{\`e}s}}, \ and\
  \bibinfo {author} {\bibfnamefont {C.-H.}\ \bibnamefont {Shen}},\ }\href
  {\doibase 10.1007/JHEP12(2025)082} {\bibfield  {journal} {\bibinfo  {journal}
  {JHEP}\ }\textbf {\bibinfo {volume} {12}},\ \bibinfo {pages} {082} (\bibinfo
  {year} {2025})},\ \Eprint {http://arxiv.org/abs/2504.18537} {arXiv:2504.18537
  [hep-ph]}\BibitemShut {NoStop}%
\bibitem [{\citenamefont {Naterop}\ and\ \citenamefont
  {Stoffer}(2025{\natexlab{b}})}]{Naterop:2025cwg}%
  \BibitemOpen
  \bibfield  {author} {\bibinfo {author} {\bibfnamefont {L.}~\bibnamefont
  {Naterop}}\ and\ \bibinfo {author} {\bibfnamefont {P.}~\bibnamefont
  {Stoffer}},\ }\href@noop {} {\  (\bibinfo {year} {2025}{\natexlab{b}})},\
  \Eprint {http://arxiv.org/abs/2507.08926} {arXiv:2507.08926 [hep-ph]}\BibitemShut {NoStop}%
\bibitem [{\citenamefont {Di~Noi}\ and\ \citenamefont
  {Gr{\"o}ber}(2025)}]{DiNoi:2025arz}%
  \BibitemOpen
  \bibfield  {author} {\bibinfo {author} {\bibfnamefont {S.}~\bibnamefont
  {Di~Noi}}\ and\ \bibinfo {author} {\bibfnamefont {R.}~\bibnamefont
  {Gr{\"o}ber}},\ }\href {\doibase 10.1016/j.physletb.2025.139878} {\bibfield
  {journal} {\bibinfo  {journal} {Phys. Lett. B}\ }\textbf {\bibinfo {volume}
  {869}},\ \bibinfo {pages} {139878} (\bibinfo {year} {2025})},\ \Eprint
  {http://arxiv.org/abs/2507.10295} {arXiv:2507.10295 [hep-ph]}\BibitemShut
  {NoStop}%
\bibitem [{\citenamefont {Haisch}\ and\ \citenamefont
  {Niggetiedt}(2025)}]{Haisch:2025vqj}%
  \BibitemOpen
  \bibfield  {author} {\bibinfo {author} {\bibfnamefont {U.}~\bibnamefont
  {Haisch}}\ and\ \bibinfo {author} {\bibfnamefont {M.}~\bibnamefont
  {Niggetiedt}},\ }\href@noop {} {\  (\bibinfo {year} {2025})},\ \Eprint
  {http://arxiv.org/abs/2507.20803} {arXiv:2507.20803 [hep-ph]}\BibitemShut
  {NoStop}%
\bibitem [{\citenamefont {Duhr}\ \emph
  {et~al.}(2025{\natexlab{b}})\citenamefont {Duhr}, \citenamefont {Ventura},\
  and\ \citenamefont {Vryonidou}}]{Duhr:2025yor}%
  \BibitemOpen
  \bibfield  {author} {\bibinfo {author} {\bibfnamefont {C.}~\bibnamefont
  {Duhr}}, \bibinfo {author} {\bibfnamefont {G.}~\bibnamefont {Ventura}}, \
  and\ \bibinfo {author} {\bibfnamefont {E.}~\bibnamefont {Vryonidou}},\ }\href
  {\doibase 10.1007/JHEP11(2025)046} {\bibfield  {journal} {\bibinfo  {journal}
  {JHEP}\ }\textbf {\bibinfo {volume} {11}},\ \bibinfo {pages} {046} (\bibinfo
  {year} {2025}{\natexlab{b}})},\ \Eprint {http://arxiv.org/abs/2508.04500}
  {arXiv:2508.04500 [hep-ph]}\BibitemShut {NoStop}%
\bibitem [{\citenamefont {Di~Noi}\ \emph
  {et~al.}(2025{\natexlab{b}})\citenamefont {Di~Noi}, \citenamefont {Erdelyi},\
  and\ \citenamefont {Gr{\"o}ber}}]{DiNoi:2025tka}%
  \BibitemOpen
  \bibfield  {author} {\bibinfo {author} {\bibfnamefont {S.}~\bibnamefont
  {Di~Noi}}, \bibinfo {author} {\bibfnamefont {B.~A.}\ \bibnamefont {Erdelyi}},
  \ and\ \bibinfo {author} {\bibfnamefont {R.}~\bibnamefont {Gr{\"o}ber}},\
  }\href@noop {} {\  (\bibinfo {year} {2025}{\natexlab{b}})},\ \Eprint
  {http://arxiv.org/abs/2510.14680} {arXiv:2510.14680 [hep-ph]}\BibitemShut
  {NoStop}%
\bibitem [{\citenamefont {Banik}\ \emph {et~al.}(2025)\citenamefont {Banik},
  \citenamefont {Crivellin}, \citenamefont {Naterop},\ and\ \citenamefont
  {Stoffer}}]{Banik:2025wpi}%
  \BibitemOpen
  \bibfield  {author} {\bibinfo {author} {\bibfnamefont {S.}~\bibnamefont
  {Banik}}, \bibinfo {author} {\bibfnamefont {A.}~\bibnamefont {Crivellin}},
  \bibinfo {author} {\bibfnamefont {L.}~\bibnamefont {Naterop}}, \ and\
  \bibinfo {author} {\bibfnamefont {P.}~\bibnamefont {Stoffer}},\ }\href@noop
  {} {\  (\bibinfo {year} {2025})},\ \Eprint {http://arxiv.org/abs/2510.08682}
  {arXiv:2510.08682 [hep-ph]}\BibitemShut {NoStop}%
\bibitem [{\citenamefont {Henriksson}\ \emph {et~al.}(2025)\citenamefont
  {Henriksson}, \citenamefont {Kousvos},\ and\ \citenamefont
  {Roosmale~Nepveu}}]{Henriksson:2025vyi}%
  \BibitemOpen
  \bibfield  {author} {\bibinfo {author} {\bibfnamefont {J.}~\bibnamefont
  {Henriksson}}, \bibinfo {author} {\bibfnamefont {S.~R.}\ \bibnamefont
  {Kousvos}}, \ and\ \bibinfo {author} {\bibfnamefont {J.}~\bibnamefont
  {Roosmale~Nepveu}},\ }\href@noop {} {\  (\bibinfo {year} {2025})},\ \Eprint
  {http://arxiv.org/abs/2511.16740} {arXiv:2511.16740 [hep-th]}\BibitemShut
  {NoStop}%
\bibitem [{\citenamefont {Chala}\ and\ \citenamefont
  {L{\'o}pez~Miras}(2025)}]{Chala:2025crd}%
  \BibitemOpen
  \bibfield  {author} {\bibinfo {author} {\bibfnamefont {M.}~\bibnamefont
  {Chala}}\ and\ \bibinfo {author} {\bibfnamefont {J.}~\bibnamefont
  {L{\'o}pez~Miras}},\ }\href@noop {} {\  (\bibinfo {year} {2025})},\ \Eprint
  {http://arxiv.org/abs/2512.04064} {arXiv:2512.04064 [hep-ph]}\BibitemShut
  {NoStop}%
\bibitem [{\citenamefont {Guedes}\ and\ \citenamefont
  {Roosmale~Nepveu}(2025)}]{Guedes:2025sax}%
  \BibitemOpen
  \bibfield  {author} {\bibinfo {author} {\bibfnamefont {G.}~\bibnamefont
  {Guedes}}\ and\ \bibinfo {author} {\bibfnamefont {J.}~\bibnamefont
  {Roosmale~Nepveu}},\ }\href@noop {} {\  (\bibinfo {year} {2025})},\ \Eprint
  {http://arxiv.org/abs/2512.08827} {arXiv:2512.08827 [hep-ph]}\BibitemShut
  {NoStop}%
\bibitem [{\citenamefont {Allwicher}\ \emph {et~al.}(2023)\citenamefont
  {Allwicher}, \citenamefont {Isidori}, \citenamefont {Lizana}, \citenamefont
  {Selimovic},\ and\ \citenamefont {Stefanek}}]{Allwicher:2023aql}%
  \BibitemOpen
  \bibfield  {author} {\bibinfo {author} {\bibfnamefont {L.}~\bibnamefont
  {Allwicher}}, \bibinfo {author} {\bibfnamefont {G.}~\bibnamefont {Isidori}},
  \bibinfo {author} {\bibfnamefont {J.~M.}\ \bibnamefont {Lizana}}, \bibinfo
  {author} {\bibfnamefont {N.}~\bibnamefont {Selimovic}}, \ and\ \bibinfo
  {author} {\bibfnamefont {B.~A.}\ \bibnamefont {Stefanek}},\ }\href {\doibase
  10.1007/JHEP05(2023)179} {\bibfield  {journal} {\bibinfo  {journal} {JHEP}\
  }\textbf {\bibinfo {volume} {05}},\ \bibinfo {pages} {179} (\bibinfo {year}
  {2023})},\ \Eprint {http://arxiv.org/abs/2302.11584} {arXiv:2302.11584
  [hep-ph]}\BibitemShut {NoStop}%
\bibitem [{\citenamefont {Allwicher}\ \emph {et~al.}(2024)\citenamefont
  {Allwicher}, \citenamefont {Cornella}, \citenamefont {Isidori},\ and\
  \citenamefont {Stefanek}}]{Allwicher:2023shc}%
  \BibitemOpen
  \bibfield  {author} {\bibinfo {author} {\bibfnamefont {L.}~\bibnamefont
  {Allwicher}}, \bibinfo {author} {\bibfnamefont {C.}~\bibnamefont {Cornella}},
  \bibinfo {author} {\bibfnamefont {G.}~\bibnamefont {Isidori}}, \ and\
  \bibinfo {author} {\bibfnamefont {B.~A.}\ \bibnamefont {Stefanek}},\ }\href
  {\doibase 10.1007/JHEP03(2024)049} {\bibfield  {journal} {\bibinfo  {journal}
  {JHEP}\ }\textbf {\bibinfo {volume} {03}},\ \bibinfo {pages} {049} (\bibinfo
  {year} {2024})},\ \Eprint {http://arxiv.org/abs/2311.00020} {arXiv:2311.00020
  [hep-ph]}\BibitemShut {NoStop}%
\bibitem [{\citenamefont {Stefanek}(2024)}]{Stefanek:2024kds}%
  \BibitemOpen
  \bibfield  {author} {\bibinfo {author} {\bibfnamefont {B.~A.}\ \bibnamefont
  {Stefanek}},\ }\href {\doibase 10.1007/JHEP09(2024)103} {\bibfield  {journal}
  {\bibinfo  {journal} {JHEP}\ }\textbf {\bibinfo {volume} {09}},\ \bibinfo
  {pages} {103} (\bibinfo {year} {2024})},\ \Eprint
  {http://arxiv.org/abs/2407.09593} {arXiv:2407.09593 [hep-ph]}\BibitemShut
  {NoStop}%
\bibitem [{\citenamefont {Haisch}\ and\ \citenamefont
  {Schnell}(2025)}]{Haisch:2024wnw}%
  \BibitemOpen
  \bibfield  {author} {\bibinfo {author} {\bibfnamefont {U.}~\bibnamefont
  {Haisch}}\ and\ \bibinfo {author} {\bibfnamefont {L.}~\bibnamefont
  {Schnell}},\ }\href {\doibase 10.1007/JHEP02(2025)038} {\bibfield  {journal}
  {\bibinfo  {journal} {JHEP}\ }\textbf {\bibinfo {volume} {02}},\ \bibinfo
  {pages} {038} (\bibinfo {year} {2025})},\ \Eprint
  {http://arxiv.org/abs/2410.13304} {arXiv:2410.13304 [hep-ph]}\BibitemShut
  {NoStop}%
\bibitem [{\citenamefont {Gauld}\ \emph {et~al.}(2016)\citenamefont {Gauld},
  \citenamefont {Pecjak},\ and\ \citenamefont {Scott}}]{Gauld:2015lmb}%
  \BibitemOpen
  \bibfield  {author} {\bibinfo {author} {\bibfnamefont {R.}~\bibnamefont
  {Gauld}}, \bibinfo {author} {\bibfnamefont {B.~D.}\ \bibnamefont {Pecjak}}, \
  and\ \bibinfo {author} {\bibfnamefont {D.~J.}\ \bibnamefont {Scott}},\ }\href
  {\doibase 10.1007/JHEP05(2016)080} {\bibfield  {journal} {\bibinfo  {journal}
  {JHEP}\ }\textbf {\bibinfo {volume} {05}},\ \bibinfo {pages} {080} (\bibinfo
  {year} {2016})},\ \Eprint {http://arxiv.org/abs/1512.02508} {arXiv:1512.02508
  [hep-ph]}\BibitemShut {NoStop}%
\bibitem [{\citenamefont {Ethier}\ \emph {et~al.}(2021)\citenamefont {Ethier},
  \citenamefont {Magni}, \citenamefont {Maltoni}, \citenamefont {Mantani},
  \citenamefont {Nocera}, \citenamefont {Rojo}, \citenamefont {Slade},
  \citenamefont {Vryonidou},\ and\ \citenamefont {Zhang}}]{Ethier:2021bye}%
  \BibitemOpen
  \bibfield  {author} {\bibinfo {author} {\bibfnamefont {J.~J.}\ \bibnamefont
  {Ethier}}, \bibinfo {author} {\bibfnamefont {G.}~\bibnamefont {Magni}},
  \bibinfo {author} {\bibfnamefont {F.}~\bibnamefont {Maltoni}}, \bibinfo
  {author} {\bibfnamefont {L.}~\bibnamefont {Mantani}}, \bibinfo {author}
  {\bibfnamefont {E.~R.}\ \bibnamefont {Nocera}}, \bibinfo {author}
  {\bibfnamefont {J.}~\bibnamefont {Rojo}}, \bibinfo {author} {\bibfnamefont
  {E.}~\bibnamefont {Slade}}, \bibinfo {author} {\bibfnamefont
  {E.}~\bibnamefont {Vryonidou}}, \ and\ \bibinfo {author} {\bibfnamefont
  {C.}~\bibnamefont {Zhang}} (\bibinfo {collaboration} {SMEFiT}),\ }\href
  {\doibase 10.1007/JHEP11(2021)089} {\bibfield  {journal} {\bibinfo  {journal}
  {JHEP}\ }\textbf {\bibinfo {volume} {11}},\ \bibinfo {pages} {089} (\bibinfo
  {year} {2021})},\ \Eprint {http://arxiv.org/abs/2105.00006} {arXiv:2105.00006
  [hep-ph]}\BibitemShut {NoStop}%
\bibitem [{\citenamefont {Dawson}\ and\ \citenamefont
  {Giardino}(2022)}]{Dawson:2022bxd}%
  \BibitemOpen
  \bibfield  {author} {\bibinfo {author} {\bibfnamefont {S.}~\bibnamefont
  {Dawson}}\ and\ \bibinfo {author} {\bibfnamefont {P.~P.}\ \bibnamefont
  {Giardino}},\ }\href {\doibase 10.1103/PhysRevD.105.073006} {\bibfield
  {journal} {\bibinfo  {journal} {Phys. Rev. D}\ }\textbf {\bibinfo {volume}
  {105}},\ \bibinfo {pages} {073006} (\bibinfo {year} {2022})},\ \Eprint
  {http://arxiv.org/abs/2201.09887} {arXiv:2201.09887 [hep-ph]}\BibitemShut
  {NoStop}%
\bibitem [{\citenamefont {Aoude}\ \emph {et~al.}(2022)\citenamefont {Aoude},
  \citenamefont {El~Faham}, \citenamefont {Maltoni},\ and\ \citenamefont
  {Vryonidou}}]{Aoude:2022deh}%
  \BibitemOpen
  \bibfield  {author} {\bibinfo {author} {\bibfnamefont {R.}~\bibnamefont
  {Aoude}}, \bibinfo {author} {\bibfnamefont {H.}~\bibnamefont {El~Faham}},
  \bibinfo {author} {\bibfnamefont {F.}~\bibnamefont {Maltoni}}, \ and\
  \bibinfo {author} {\bibfnamefont {E.}~\bibnamefont {Vryonidou}},\ }\href
  {\doibase 10.1007/JHEP10(2022)163} {\bibfield  {journal} {\bibinfo  {journal}
  {JHEP}\ }\textbf {\bibinfo {volume} {10}},\ \bibinfo {pages} {163} (\bibinfo
  {year} {2022})},\ \Eprint {http://arxiv.org/abs/2208.04962} {arXiv:2208.04962
  [hep-ph]}\BibitemShut {NoStop}%
\bibitem [{\citenamefont {Aad}\ \emph {et~al.}(2023)\citenamefont {Aad} \emph
  {et~al.}}]{ATLAS:2023ajo}%
  \BibitemOpen
  \bibfield  {author} {\bibinfo {author} {\bibfnamefont {G.}~\bibnamefont
  {Aad}} \emph {et~al.} (\bibinfo {collaboration} {ATLAS}),\ }\href {\doibase
  10.1140/epjc/s10052-023-11573-0} {\bibfield  {journal} {\bibinfo  {journal}
  {Eur. Phys. J. C}\ }\textbf {\bibinfo {volume} {83}},\ \bibinfo {pages} {496}
  (\bibinfo {year} {2023})},\ \Eprint {http://arxiv.org/abs/2303.15061}
  {arXiv:2303.15061 [hep-ex]}\BibitemShut {NoStop}%
\bibitem [{\citenamefont {Degrande}\ \emph {et~al.}(2024)\citenamefont
  {Degrande}, \citenamefont {Rosenfeld},\ and\ \citenamefont
  {Vasquez}}]{Degrande:2024mbg}%
  \BibitemOpen
  \bibfield  {author} {\bibinfo {author} {\bibfnamefont {C.}~\bibnamefont
  {Degrande}}, \bibinfo {author} {\bibfnamefont {R.}~\bibnamefont {Rosenfeld}},
  \ and\ \bibinfo {author} {\bibfnamefont {A.}~\bibnamefont {Vasquez}},\ }\href
  {\doibase 10.1007/JHEP07(2024)114} {\bibfield  {journal} {\bibinfo  {journal}
  {JHEP}\ }\textbf {\bibinfo {volume} {07}},\ \bibinfo {pages} {114} (\bibinfo
  {year} {2024})},\ \Eprint {http://arxiv.org/abs/2402.06528} {arXiv:2402.06528
  [hep-ph]}\BibitemShut {NoStop}%
\bibitem [{\citenamefont {Di~Noi}\ \emph {et~al.}(2026)\citenamefont {Di~Noi},
  \citenamefont {El~Faham}, \citenamefont {Gr{\"o}ber}, \citenamefont {Vitti},\
  and\ \citenamefont {Vryonidou}}]{DiNoi:2025uhu}%
  \BibitemOpen
  \bibfield  {author} {\bibinfo {author} {\bibfnamefont {S.}~\bibnamefont
  {Di~Noi}}, \bibinfo {author} {\bibfnamefont {H.}~\bibnamefont {El~Faham}},
  \bibinfo {author} {\bibfnamefont {R.}~\bibnamefont {Gr{\"o}ber}}, \bibinfo
  {author} {\bibfnamefont {M.}~\bibnamefont {Vitti}}, \ and\ \bibinfo {author}
  {\bibfnamefont {E.}~\bibnamefont {Vryonidou}},\ }\href {\doibase
  10.1007/JHEP01(2026)025} {\bibfield  {journal} {\bibinfo  {journal} {JHEP}\
  }\textbf {\bibinfo {volume} {01}},\ \bibinfo {pages} {025} (\bibinfo {year}
  {2026})},\ \Eprint {http://arxiv.org/abs/2507.01137} {arXiv:2507.01137
  [hep-ph]}\BibitemShut {NoStop}%
\bibitem [{\citenamefont {Krauss}\ \emph {et~al.}(2017)\citenamefont {Krauss},
  \citenamefont {Kuttimalai},\ and\ \citenamefont {Plehn}}]{Krauss:2016ely}%
  \BibitemOpen
  \bibfield  {author} {\bibinfo {author} {\bibfnamefont {F.}~\bibnamefont
  {Krauss}}, \bibinfo {author} {\bibfnamefont {S.}~\bibnamefont {Kuttimalai}},
  \ and\ \bibinfo {author} {\bibfnamefont {T.}~\bibnamefont {Plehn}},\ }\href
  {\doibase 10.1103/PhysRevD.95.035024} {\bibfield  {journal} {\bibinfo
  {journal} {Phys. Rev. D}\ }\textbf {\bibinfo {volume} {95}},\ \bibinfo
  {pages} {035024} (\bibinfo {year} {2017})},\ \Eprint
  {http://arxiv.org/abs/1611.00767} {arXiv:1611.00767 [hep-ph]}\BibitemShut
  {NoStop}%
\bibitem [{\citenamefont {Alioli}\ \emph {et~al.}(2017)\citenamefont {Alioli},
  \citenamefont {Farina}, \citenamefont {Pappadopulo},\ and\ \citenamefont
  {Ruderman}}]{Alioli:2017jdo}%
  \BibitemOpen
  \bibfield  {author} {\bibinfo {author} {\bibfnamefont {S.}~\bibnamefont
  {Alioli}}, \bibinfo {author} {\bibfnamefont {M.}~\bibnamefont {Farina}},
  \bibinfo {author} {\bibfnamefont {D.}~\bibnamefont {Pappadopulo}}, \ and\
  \bibinfo {author} {\bibfnamefont {J.~T.}\ \bibnamefont {Ruderman}},\ }\href
  {\doibase 10.1007/JHEP07(2017)097} {\bibfield  {journal} {\bibinfo  {journal}
  {JHEP}\ }\textbf {\bibinfo {volume} {07}},\ \bibinfo {pages} {097} (\bibinfo
  {year} {2017})},\ \Eprint {http://arxiv.org/abs/1706.03068} {arXiv:1706.03068
  [hep-ph]}\BibitemShut {NoStop}%
\bibitem [{\citenamefont {Alte}\ \emph {et~al.}(2018)\citenamefont {Alte},
  \citenamefont {K{\"o}nig},\ and\ \citenamefont {Shepherd}}]{Alte:2017pme}%
  \BibitemOpen
  \bibfield  {author} {\bibinfo {author} {\bibfnamefont {S.}~\bibnamefont
  {Alte}}, \bibinfo {author} {\bibfnamefont {M.}~\bibnamefont {K{\"o}nig}}, \
  and\ \bibinfo {author} {\bibfnamefont {W.}~\bibnamefont {Shepherd}},\ }\href
  {\doibase 10.1007/JHEP01(2018)094} {\bibfield  {journal} {\bibinfo  {journal}
  {JHEP}\ }\textbf {\bibinfo {volume} {01}},\ \bibinfo {pages} {094} (\bibinfo
  {year} {2018})},\ \Eprint {http://arxiv.org/abs/1711.07484} {arXiv:1711.07484
  [hep-ph]}\BibitemShut {NoStop}%
\bibitem [{\citenamefont {Hirschi}\ \emph {et~al.}(2018)\citenamefont
  {Hirschi}, \citenamefont {Maltoni}, \citenamefont {Tsinikos},\ and\
  \citenamefont {Vryonidou}}]{Hirschi:2018etq}%
  \BibitemOpen
  \bibfield  {author} {\bibinfo {author} {\bibfnamefont {V.}~\bibnamefont
  {Hirschi}}, \bibinfo {author} {\bibfnamefont {F.}~\bibnamefont {Maltoni}},
  \bibinfo {author} {\bibfnamefont {I.}~\bibnamefont {Tsinikos}}, \ and\
  \bibinfo {author} {\bibfnamefont {E.}~\bibnamefont {Vryonidou}},\ }\href
  {\doibase 10.1007/JHEP07(2018)093} {\bibfield  {journal} {\bibinfo  {journal}
  {JHEP}\ }\textbf {\bibinfo {volume} {07}},\ \bibinfo {pages} {093} (\bibinfo
  {year} {2018})},\ \Eprint {http://arxiv.org/abs/1806.04696} {arXiv:1806.04696
  [hep-ph]}\BibitemShut {NoStop}%
\bibitem [{\citenamefont {Keilmann}\ and\ \citenamefont
  {Shepherd}(2019)}]{Keilmann:2019cbp}%
  \BibitemOpen
  \bibfield  {author} {\bibinfo {author} {\bibfnamefont {E.}~\bibnamefont
  {Keilmann}}\ and\ \bibinfo {author} {\bibfnamefont {W.}~\bibnamefont
  {Shepherd}},\ }\href {\doibase 10.1007/JHEP09(2019)086} {\bibfield  {journal}
  {\bibinfo  {journal} {JHEP}\ }\textbf {\bibinfo {volume} {09}},\ \bibinfo
  {pages} {086} (\bibinfo {year} {2019})},\ \Eprint
  {http://arxiv.org/abs/1907.13160} {arXiv:1907.13160 [hep-ph]}\BibitemShut
  {NoStop}%
\bibitem [{\citenamefont {Goldouzian}\ and\ \citenamefont
  {Hildreth}(2020)}]{Goldouzian:2020wdq}%
  \BibitemOpen
  \bibfield  {author} {\bibinfo {author} {\bibfnamefont {R.}~\bibnamefont
  {Goldouzian}}\ and\ \bibinfo {author} {\bibfnamefont {M.~D.}\ \bibnamefont
  {Hildreth}},\ }\href {\doibase 10.1016/j.physletb.2020.135889} {\bibfield
  {journal} {\bibinfo  {journal} {Phys. Lett. B}\ }\textbf {\bibinfo {volume}
  {811}},\ \bibinfo {pages} {135889} (\bibinfo {year} {2020})},\ \Eprint
  {http://arxiv.org/abs/2001.02736} {arXiv:2001.02736 [hep-ph]}\BibitemShut
  {NoStop}%
\bibitem [{\citenamefont {Haisch}\ and\ \citenamefont
  {Koole}(2021)}]{Haisch:2021hcg}%
  \BibitemOpen
  \bibfield  {author} {\bibinfo {author} {\bibfnamefont {U.}~\bibnamefont
  {Haisch}}\ and\ \bibinfo {author} {\bibfnamefont {G.}~\bibnamefont {Koole}},\
  }\href {\doibase 10.1007/JHEP09(2021)133} {\bibfield  {journal} {\bibinfo
  {journal} {JHEP}\ }\textbf {\bibinfo {volume} {09}},\ \bibinfo {pages} {133}
  (\bibinfo {year} {2021})},\ \Eprint {http://arxiv.org/abs/2106.01289}
  {arXiv:2106.01289 [hep-ph]}\BibitemShut {NoStop}%
\bibitem [{\citenamefont {Degrande}\ and\ \citenamefont
  {Maltoni}(2025)}]{Degrande:2025vhl}%
  \BibitemOpen
  \bibfield  {author} {\bibinfo {author} {\bibfnamefont {C.}~\bibnamefont
  {Degrande}}\ and\ \bibinfo {author} {\bibfnamefont {M.}~\bibnamefont
  {Maltoni}},\ }\href@noop {} {\  (\bibinfo {year} {2025})},\ \Eprint
  {http://arxiv.org/abs/2511.04517} {arXiv:2511.04517 [hep-ph]}\BibitemShut
  {NoStop}%
\bibitem [{\citenamefont {Buras}\ and\ \citenamefont
  {Jung}(2018)}]{Buras:2018gto}%
  \BibitemOpen
  \bibfield  {author} {\bibinfo {author} {\bibfnamefont {A.~J.}\ \bibnamefont
  {Buras}}\ and\ \bibinfo {author} {\bibfnamefont {M.}~\bibnamefont {Jung}},\
  }\href {\doibase 10.1007/JHEP06(2018)067} {\bibfield  {journal} {\bibinfo
  {journal} {JHEP}\ }\textbf {\bibinfo {volume} {06}},\ \bibinfo {pages} {067}
  (\bibinfo {year} {2018})},\ \Eprint {http://arxiv.org/abs/1804.05852}
  {arXiv:1804.05852 [hep-ph]}\BibitemShut {NoStop}%
\bibitem [{\citenamefont {Sirunyan}\ \emph {et~al.}(2018)\citenamefont
  {Sirunyan} \emph {et~al.}}]{CMS:2018ucw}%
  \BibitemOpen
  \bibfield  {author} {\bibinfo {author} {\bibfnamefont {A.~M.}\ \bibnamefont
  {Sirunyan}} \emph {et~al.} (\bibinfo {collaboration} {CMS}),\ }\href
  {\doibase 10.1140/epjc/s10052-018-6242-x} {\bibfield  {journal} {\bibinfo
  {journal} {Eur. Phys. J. C}\ }\textbf {\bibinfo {volume} {78}},\ \bibinfo
  {pages} {789} (\bibinfo {year} {2018})},\ \bibinfo {note} {[Erratum: Eur.
  Phys. J. C {\bf 82}, 379 (2022)]},\ \Eprint {http://arxiv.org/abs/1803.08030}
  {arXiv:1803.08030 [hep-ex]}\BibitemShut {NoStop}%
\bibitem [{\citenamefont {Hayrapetyan}\ \emph {et~al.}(2025)\citenamefont
  {Hayrapetyan} \emph {et~al.}}]{CMS:2023fix}%
  \BibitemOpen
  \bibfield  {author} {\bibinfo {author} {\bibfnamefont {A.}~\bibnamefont
  {Hayrapetyan}} \emph {et~al.} (\bibinfo {collaboration} {CMS}),\ }\href
  {\doibase 10.1140/epjc/s10052-024-13606-8} {\bibfield  {journal} {\bibinfo
  {journal} {Eur. Phys. J. C}\ }\textbf {\bibinfo {volume} {85}},\ \bibinfo
  {pages} {72} (\bibinfo {year} {2025})},\ \Eprint
  {http://arxiv.org/abs/2312.16669} {arXiv:2312.16669 [hep-ex]}\BibitemShut
  {NoStop}%
\bibitem [{\citenamefont {Aad}\ \emph {et~al.}(2025)\citenamefont {Aad} \emph
  {et~al.}}]{ATLAS:2025ifq}%
  \BibitemOpen
  \bibfield  {author} {\bibinfo {author} {\bibfnamefont {G.}~\bibnamefont
  {Aad}} \emph {et~al.} (\bibinfo {collaboration} {ATLAS}),\ }\href@noop {} {\
  (\bibinfo {year} {2025})},\ \Eprint {http://arxiv.org/abs/2512.19073}
  {arXiv:2512.19073 [hep-ex]}\BibitemShut {NoStop}%
\bibitem [{\citenamefont {Dulat}\ \emph {et~al.}(2016)\citenamefont {Dulat},
  \citenamefont {Hou}, \citenamefont {Gao}, \citenamefont {Guzzi},
  \citenamefont {Huston}, \citenamefont {Nadolsky}, \citenamefont {Pumplin},
  \citenamefont {Schmidt}, \citenamefont {Stump},\ and\ \citenamefont
  {Yuan}}]{Dulat:2015mca}%
  \BibitemOpen
  \bibfield  {author} {\bibinfo {author} {\bibfnamefont {S.}~\bibnamefont
  {Dulat}}, \bibinfo {author} {\bibfnamefont {T.-J.}\ \bibnamefont {Hou}},
  \bibinfo {author} {\bibfnamefont {J.}~\bibnamefont {Gao}}, \bibinfo {author}
  {\bibfnamefont {M.}~\bibnamefont {Guzzi}}, \bibinfo {author} {\bibfnamefont
  {J.}~\bibnamefont {Huston}}, \bibinfo {author} {\bibfnamefont
  {P.}~\bibnamefont {Nadolsky}}, \bibinfo {author} {\bibfnamefont
  {J.}~\bibnamefont {Pumplin}}, \bibinfo {author} {\bibfnamefont
  {C.}~\bibnamefont {Schmidt}}, \bibinfo {author} {\bibfnamefont
  {D.}~\bibnamefont {Stump}}, \ and\ \bibinfo {author} {\bibfnamefont {C.~P.}\
  \bibnamefont {Yuan}},\ }\href {\doibase 10.1103/PhysRevD.93.033006}
  {\bibfield  {journal} {\bibinfo  {journal} {Phys. Rev. D}\ }\textbf {\bibinfo
  {volume} {93}},\ \bibinfo {pages} {033006} (\bibinfo {year} {2016})},\
  \Eprint {http://arxiv.org/abs/1506.07443} {arXiv:1506.07443 [hep-ph]}
\BibitemShut {NoStop}%
\bibitem [{\citenamefont {Clark}\ \emph {et~al.}(2017)\citenamefont {Clark},
  \citenamefont {Godat},\ and\ \citenamefont {Olness}}]{Clark:2016jgm}%
  \BibitemOpen
  \bibfield  {author} {\bibinfo {author} {\bibfnamefont {D.~B.}\ \bibnamefont
  {Clark}}, \bibinfo {author} {\bibfnamefont {E.}~\bibnamefont {Godat}}, \ and\
  \bibinfo {author} {\bibfnamefont {F.~I.}\ \bibnamefont {Olness}},\ }\href
  {\doibase 10.1016/j.cpc.2017.03.004} {\bibfield  {journal} {\bibinfo
  {journal} {Comput. Phys. Commun.}\ }\textbf {\bibinfo {volume} {216}},\
  \bibinfo {pages} {126} (\bibinfo {year} {2017})},\ \Eprint
  {http://arxiv.org/abs/1605.08012} {arXiv:1605.08012 [hep-ph]}\BibitemShut
  {NoStop}%
\bibitem [{\citenamefont {Nagy}(2002)}]{Nagy:2001fj}%
  \BibitemOpen
  \bibfield  {author} {\bibinfo {author} {\bibfnamefont {Z.}~\bibnamefont
  {Nagy}},\ }\href {\doibase 10.1103/PhysRevLett.88.122003} {\bibfield
  {journal} {\bibinfo  {journal} {Phys. Rev. Lett.}\ }\textbf {\bibinfo
  {volume} {88}},\ \bibinfo {pages} {122003} (\bibinfo {year} {2002})},\
  \Eprint {http://arxiv.org/abs/hep-ph/0110315} {arXiv:hep-ph/0110315}\BibitemShut {NoStop}%
\bibitem [{\citenamefont {Dittmaier}\ \emph {et~al.}(2012)\citenamefont
  {Dittmaier}, \citenamefont {Huss},\ and\ \citenamefont
  {Speckner}}]{Dittmaier:2012kx}%
  \BibitemOpen
  \bibfield  {author} {\bibinfo {author} {\bibfnamefont {S.}~\bibnamefont
  {Dittmaier}}, \bibinfo {author} {\bibfnamefont {A.}~\bibnamefont {Huss}}, \
  and\ \bibinfo {author} {\bibfnamefont {C.}~\bibnamefont {Speckner}},\ }\href
  {\doibase 10.1007/JHEP11(2012)095} {\bibfield  {journal} {\bibinfo  {journal}
  {JHEP}\ }\textbf {\bibinfo {volume} {11}},\ \bibinfo {pages} {095} (\bibinfo
  {year} {2012})},\ \Eprint {http://arxiv.org/abs/1210.0438} {arXiv:1210.0438
  [hep-ph]}\BibitemShut {NoStop}%
\bibitem [{\citenamefont {Sjostrand}\ \emph {et~al.}(2008)\citenamefont
  {Sjostrand}, \citenamefont {Mrenna},\ and\ \citenamefont
  {Skands}}]{Sjostrand:2007gs}%
  \BibitemOpen
  \bibfield  {author} {\bibinfo {author} {\bibfnamefont {T.}~\bibnamefont
  {Sjostrand}}, \bibinfo {author} {\bibfnamefont {S.}~\bibnamefont {Mrenna}}, \
  and\ \bibinfo {author} {\bibfnamefont {P.~Z.}\ \bibnamefont {Skands}},\
  }\href {\doibase 10.1016/j.cpc.2008.01.036} {\bibfield  {journal} {\bibinfo
  {journal} {Comput. Phys. Commun.}\ }\textbf {\bibinfo {volume} {178}},\
  \bibinfo {pages} {852} (\bibinfo {year} {2008})},\ \Eprint
  {http://arxiv.org/abs/0710.3820} {arXiv:0710.3820 [hep-ph]}\BibitemShut
  {NoStop}%
\bibitem [{\citenamefont {Bahr}\ \emph {et~al.}(2008)\citenamefont {Bahr} \emph
  {et~al.}}]{Bahr:2008pv}%
  \BibitemOpen
  \bibfield  {author} {\bibinfo {author} {\bibfnamefont {M.}~\bibnamefont
  {Bahr}} \emph {et~al.},\ }\href {\doibase 10.1140/epjc/s10052-008-0798-9}
  {\bibfield  {journal} {\bibinfo  {journal} {Eur. Phys. J. C}\ }\textbf
  {\bibinfo {volume} {58}},\ \bibinfo {pages} {639} (\bibinfo {year} {2008})},\
  \Eprint {http://arxiv.org/abs/0803.0883} {arXiv:0803.0883 [hep-ph]}\BibitemShut {NoStop}%
\bibitem [{\citenamefont {Chen}\ \emph {et~al.}(2022)\citenamefont {Chen},
  \citenamefont {Gehrmann}, \citenamefont {Glover}, \citenamefont {Huss},\ and\
  \citenamefont {Mo}}]{Chen:2022tpk}%
  \BibitemOpen
  \bibfield  {author} {\bibinfo {author} {\bibfnamefont {X.}~\bibnamefont
  {Chen}}, \bibinfo {author} {\bibfnamefont {T.}~\bibnamefont {Gehrmann}},
  \bibinfo {author} {\bibfnamefont {E.~W.~N.}\ \bibnamefont {Glover}}, \bibinfo
  {author} {\bibfnamefont {A.}~\bibnamefont {Huss}}, \ and\ \bibinfo {author}
  {\bibfnamefont {J.}~\bibnamefont {Mo}},\ }\href {\doibase
  10.1007/JHEP09(2022)025} {\bibfield  {journal} {\bibinfo  {journal} {JHEP}\
  }\textbf {\bibinfo {volume} {09}},\ \bibinfo {pages} {025} (\bibinfo {year}
  {2022})},\ \Eprint {http://arxiv.org/abs/2204.10173} {arXiv:2204.10173
  [hep-ph]}\BibitemShut {NoStop}%
\bibitem [{\citenamefont {Huss}\ \emph {et~al.}(2025)\citenamefont {Huss} \emph
  {et~al.}}]{NNLOJET:2025rno}%
  \BibitemOpen
  \bibfield  {author} {\bibinfo {author} {\bibfnamefont {A.}~\bibnamefont
  {Huss}} \emph {et~al.} (\bibinfo {collaboration} {NNLOJET}),\ }\href@noop {}
  {\  (\bibinfo {year} {2025})},\ \Eprint {http://arxiv.org/abs/2503.22804}
  {arXiv:2503.22804 [hep-ph]}\BibitemShut {NoStop}%
\bibitem [{\citenamefont {Aaboud}\ \emph {et~al.}(2019)\citenamefont {Aaboud}
  \emph {et~al.}}]{ATLAS:2018fwl}%
  \BibitemOpen
  \bibfield  {author} {\bibinfo {author} {\bibfnamefont {M.}~\bibnamefont
  {Aaboud}} \emph {et~al.} (\bibinfo {collaboration} {ATLAS}),\ }\href
  {\doibase 10.1007/JHEP04(2019)046} {\bibfield  {journal} {\bibinfo  {journal}
  {JHEP}\ }\textbf {\bibinfo {volume} {04}},\ \bibinfo {pages} {046} (\bibinfo
  {year} {2019})},\ \Eprint {http://arxiv.org/abs/1811.12113} {arXiv:1811.12113
  [hep-ex]}\BibitemShut {NoStop}%
\bibitem [{\citenamefont {Sirunyan}\ \emph
  {et~al.}(2020{\natexlab{a}})\citenamefont {Sirunyan} \emph
  {et~al.}}]{CMS:2019rvj}%
  \BibitemOpen
  \bibfield  {author} {\bibinfo {author} {\bibfnamefont {A.~M.}\ \bibnamefont
  {Sirunyan}} \emph {et~al.} (\bibinfo {collaboration} {CMS}),\ }\href
  {\doibase 10.1140/epjc/s10052-019-7593-7} {\bibfield  {journal} {\bibinfo
  {journal} {Eur. Phys. J. C}\ }\textbf {\bibinfo {volume} {80}},\ \bibinfo
  {pages} {75} (\bibinfo {year} {2020}{\natexlab{a}})},\ \Eprint
  {http://arxiv.org/abs/1908.06463} {arXiv:1908.06463 [hep-ex]}\BibitemShut
  {NoStop}%
\bibitem [{\citenamefont {Sirunyan}\ \emph
  {et~al.}(2020{\natexlab{b}})\citenamefont {Sirunyan} \emph
  {et~al.}}]{CMS:2019eih}%
  \BibitemOpen
  \bibfield  {author} {\bibinfo {author} {\bibfnamefont {A.~M.}\ \bibnamefont
  {Sirunyan}} \emph {et~al.} (\bibinfo {collaboration} {CMS}),\ }\href
  {\doibase 10.1016/j.physletb.2020.135285} {\bibfield  {journal} {\bibinfo
  {journal} {Phys. Lett. B}\ }\textbf {\bibinfo {volume} {803}},\ \bibinfo
  {pages} {135285} (\bibinfo {year} {2020}{\natexlab{b}})},\ \Eprint
  {http://arxiv.org/abs/1909.05306} {arXiv:1909.05306 [hep-ex]}\BibitemShut
  {NoStop}%
\bibitem [{\citenamefont {Aad}\ \emph {et~al.}(2020)\citenamefont {Aad} \emph
  {et~al.}}]{ATLAS:2020hpj}%
  \BibitemOpen
  \bibfield  {author} {\bibinfo {author} {\bibfnamefont {G.}~\bibnamefont
  {Aad}} \emph {et~al.} (\bibinfo {collaboration} {ATLAS}),\ }\href {\doibase
  10.1140/epjc/s10052-020-08509-3} {\bibfield  {journal} {\bibinfo  {journal}
  {Eur. Phys. J. C}\ }\textbf {\bibinfo {volume} {80}},\ \bibinfo {pages}
  {1085} (\bibinfo {year} {2020})},\ \Eprint {http://arxiv.org/abs/2007.14858}
  {arXiv:2007.14858 [hep-ex]}\BibitemShut {NoStop}%
\bibitem [{\citenamefont {Banelli}\ \emph {et~al.}(2021)\citenamefont
  {Banelli}, \citenamefont {Salvioni}, \citenamefont {Serra}, \citenamefont
  {Theil},\ and\ \citenamefont {Weiler}}]{Banelli:2020iau}%
  \BibitemOpen
  \bibfield  {author} {\bibinfo {author} {\bibfnamefont {G.}~\bibnamefont
  {Banelli}}, \bibinfo {author} {\bibfnamefont {E.}~\bibnamefont {Salvioni}},
  \bibinfo {author} {\bibfnamefont {J.}~\bibnamefont {Serra}}, \bibinfo
  {author} {\bibfnamefont {T.}~\bibnamefont {Theil}}, \ and\ \bibinfo {author}
  {\bibfnamefont {A.}~\bibnamefont {Weiler}},\ }\href {\doibase
  10.1007/JHEP02(2021)043} {\bibfield  {journal} {\bibinfo  {journal} {JHEP}\
  }\textbf {\bibinfo {volume} {02}},\ \bibinfo {pages} {043} (\bibinfo {year}
  {2021})},\ \Eprint {http://arxiv.org/abs/2010.05915} {arXiv:2010.05915
  [hep-ph]}\BibitemShut {NoStop}%
\bibitem [{\citenamefont {Aad}\ \emph {et~al.}(2021)\citenamefont {Aad} \emph
  {et~al.}}]{ATLAS:2021kqb}%
  \BibitemOpen
  \bibfield  {author} {\bibinfo {author} {\bibfnamefont {G.}~\bibnamefont
  {Aad}} \emph {et~al.} (\bibinfo {collaboration} {ATLAS}),\ }\href {\doibase
  10.1007/JHEP11(2021)118} {\bibfield  {journal} {\bibinfo  {journal} {JHEP}\
  }\textbf {\bibinfo {volume} {11}},\ \bibinfo {pages} {118} (\bibinfo {year}
  {2021})},\ \Eprint {http://arxiv.org/abs/2106.11683} {arXiv:2106.11683
  [hep-ex]}\BibitemShut {NoStop}%
\bibitem [{\citenamefont {Tumasyan}\ \emph {et~al.}(2023)\citenamefont
  {Tumasyan} \emph {et~al.}}]{CMS:2023zdh}%
  \BibitemOpen
  \bibfield  {author} {\bibinfo {author} {\bibfnamefont {A.}~\bibnamefont
  {Tumasyan}} \emph {et~al.} (\bibinfo {collaboration} {CMS}),\ }\href
  {\doibase 10.1016/j.physletb.2023.138076} {\bibfield  {journal} {\bibinfo
  {journal} {Phys. Lett. B}\ }\textbf {\bibinfo {volume} {844}},\ \bibinfo
  {pages} {138076} (\bibinfo {year} {2023})},\ \Eprint
  {http://arxiv.org/abs/2303.03864} {arXiv:2303.03864 [hep-ex]}\BibitemShut
  {NoStop}%
\bibitem [{CMS(2025)}]{CMS-PAS-TOP-24-008}%
  \BibitemOpen
  \href {https://cds.cern.ch/record/2944177} {\emph {\bibinfo {title} {{Search
  for physics beyond the standard model in four and three top quark production
  events using proton-proton collisions at $\sqrt{s}=13\,\mathrm{TeV}$}}}},\
  \bibinfo {type} {Tech. Rep.}\ (\bibinfo  {institution} {CERN},\ \bibinfo
  {address} {Geneva},\ \bibinfo {year} {2025})\BibitemShut {NoStop}%
\bibitem [{\citenamefont {Alloul}\ \emph {et~al.}(2014)\citenamefont {Alloul},
  \citenamefont {Christensen}, \citenamefont {Degrande}, \citenamefont {Duhr},\
  and\ \citenamefont {Fuks}}]{Alloul:2013bka}%
  \BibitemOpen
  \bibfield  {author} {\bibinfo {author} {\bibfnamefont {A.}~\bibnamefont
  {Alloul}}, \bibinfo {author} {\bibfnamefont {N.~D.}\ \bibnamefont
  {Christensen}}, \bibinfo {author} {\bibfnamefont {C.}~\bibnamefont
  {Degrande}}, \bibinfo {author} {\bibfnamefont {C.}~\bibnamefont {Duhr}}, \
  and\ \bibinfo {author} {\bibfnamefont {B.}~\bibnamefont {Fuks}},\ }\href
  {\doibase 10.1016/j.cpc.2014.04.012} {\bibfield  {journal} {\bibinfo
  {journal} {Comput. Phys. Commun.}\ }\textbf {\bibinfo {volume} {185}},\
  \bibinfo {pages} {2250} (\bibinfo {year} {2014})},\ \Eprint
  {http://arxiv.org/abs/1310.1921} {arXiv:1310.1921 [hep-ph]}\BibitemShut
  {NoStop}%
\bibitem [{\citenamefont {Hahn}(2001)}]{Hahn:2000kx}%
  \BibitemOpen
  \bibfield  {author} {\bibinfo {author} {\bibfnamefont {T.}~\bibnamefont
  {Hahn}},\ }\href {\doibase 10.1016/S0010-4655(01)00290-9} {\bibfield
  {journal} {\bibinfo  {journal} {Comput. Phys. Commun.}\ }\textbf {\bibinfo
  {volume} {140}},\ \bibinfo {pages} {418} (\bibinfo {year} {2001})},\ \Eprint
  {http://arxiv.org/abs/hep-ph/0012260} {arXiv:hep-ph/0012260}\BibitemShut
  {NoStop}%
\bibitem [{\citenamefont {Hahn}\ \emph {et~al.}(2016)\citenamefont {Hahn},
  \citenamefont {Pa\ss{}ehr},\ and\ \citenamefont
  {Schappacher}}]{Hahn:2016ebn}%
  \BibitemOpen
  \bibfield  {author} {\bibinfo {author} {\bibfnamefont {T.}~\bibnamefont
  {Hahn}}, \bibinfo {author} {\bibfnamefont {S.}~\bibnamefont {Pa\ss{}ehr}}, \
  and\ \bibinfo {author} {\bibfnamefont {C.}~\bibnamefont {Schappacher}},\
  }\href {\doibase 10.1088/1742-6596/762/1/012065} {\bibfield  {journal}
  {\bibinfo  {journal} {PoS}\ }\textbf {\bibinfo {volume} {LL2016}},\ \bibinfo
  {pages} {068} (\bibinfo {year} {2016})},\ \Eprint
  {http://arxiv.org/abs/1604.04611} {arXiv:1604.04611 [hep-ph]}\BibitemShut
  {NoStop}%
\bibitem [{\citenamefont {Fuentes-Martin}\ \emph {et~al.}(2021)\citenamefont
  {Fuentes-Martin}, \citenamefont {Ruiz-Femenia}, \citenamefont {Vicente},\
  and\ \citenamefont {Virto}}]{Fuentes-Martin:2020zaz}%
  \BibitemOpen
  \bibfield  {author} {\bibinfo {author} {\bibfnamefont {J.}~\bibnamefont
  {Fuentes-Martin}}, \bibinfo {author} {\bibfnamefont {P.}~\bibnamefont
  {Ruiz-Femenia}}, \bibinfo {author} {\bibfnamefont {A.}~\bibnamefont
  {Vicente}}, \ and\ \bibinfo {author} {\bibfnamefont {J.}~\bibnamefont
  {Virto}},\ }\href {\doibase 10.1140/epjc/s10052-020-08778-y} {\bibfield
  {journal} {\bibinfo  {journal} {Eur. Phys. J. C}\ }\textbf {\bibinfo {volume}
  {81}},\ \bibinfo {pages} {167} (\bibinfo {year} {2021})},\ \Eprint
  {http://arxiv.org/abs/2010.16341} {arXiv:2010.16341 [hep-ph]}\BibitemShut
  {NoStop}%
\bibitem [{\citenamefont {Barducci}\ \emph {et~al.}(2018)\citenamefont
  {Barducci} \emph {et~al.}}]{Aguilar-Saavedra:2018ksv}%
  \BibitemOpen
  \bibfield  {author} {\bibinfo {author} {\bibfnamefont {D.}~\bibnamefont
  {Barducci}} \emph {et~al.},\ }\href@noop {} {\  (\bibinfo {year} {2018})},\
  \Eprint {http://arxiv.org/abs/1802.07237} {arXiv:1802.07237 [hep-ph]}\BibitemShut {NoStop}%
\end{thebibliography}


%

\end{document}